\documentclass[final,12pt]{elsarticle}

\usepackage{amssymb}
\usepackage{amsmath}
\usepackage{lineno}
\usepackage{xcolor}
\usepackage[dvipsnames]{xcolor}

\usepackage{subcaption}
\usepackage{siunitx}

\usepackage{tikz}
\usepackage{amsmath, amssymb, latexsym}

\usetikzlibrary{shapes, shadows, arrows}
\usetikzlibrary{positioning}
\usetikzlibrary{shapes.geometric, arrows, calc, fit, positioning, chains, arrows.meta}
\usetikzlibrary{decorations.pathreplacing}
\usetikzlibrary{fadings}

\usepackage{tabularx}
\usepackage{booktabs}
\usepackage{multirow}
\usepackage{makecell}
\usepackage{libertine}

\usepackage{tablefootnote}
\usepackage{float} 

\usepackage{microtype}
\usepackage{adjustbox}
\usepackage{ragged2e}
\usepackage[font=small,labelfont=bf]{caption}
\usepackage{placeins}

\usepackage{soul}

\begin{document}

\begin{frontmatter}

\title{A GAN-Based Framework for Generating STFT Spectrograms of Rare Acoustic Events in Structural Health Monitoring}

\author[label1,label2]{Sasan Farhadi}
\affiliation[label1]{organization={Department of Structural, Geotechnical and Building Engineering, Politecnico di Torino}, 
city={Torino},
country={Italy}}
\affiliation[label3]{organization={Department of Civil, Environmental and Geomatic Engineering, ETH Zürich
}, 
city={Zurich},
country={Switzerland}}    
\affiliation[label2]{organization={Institute of Engineering Geodesy and Measurement Systems, Graz University of Technology}, 
city={Graz},
country={Austria}}

\author[label1]{Mariateresa Iavarone}
\author[label1]{Mauro Corrado}
\author[label3]{Eleni Chatzi}
\author[label1]{Giulio Ventura}

\begin{abstract}
Structural Health Monitoring plays a crucial role in ensuring the safety, reliability, and longevity of bridge infrastructures through early damage detection. Although recent advances in deep learning-based models have enabled automated event detection, their performance is often limited by data scarcity, environmental noise, and class imbalance. To address these challenges, this study introduces a customized Generative Adversarial Network model, STFTSynth, designed particularly for generating short-time Fourier transform spectrograms derived from acoustic event signals. In contrast to augmentation techniques such as MixUp, generative adversarial networks can synthesize spectrograms that visually and statistically resemble real event representations, providing a basis for representation-level dataset enrichment. The proposed model integrates dense residual blocks for spatial consistency with bidirectional gated recurrent units for temporal dependency modeling. Model performance is evaluated against three baseline generative models using qualitative inspection and quantitative metrics, including Structural Similarity Index Measure, Peak Signal-to-Noise Ratio, and Fréchet Inception Distance. Results show that STFTSynth outperforms baseline models, producing high-resolution, temporally consistent spectrograms that align closely with real-world data. These findings highlight the potential of GAN-based spectrogram synthesis for representation-level enrichment of rare acoustic-event datasets in bridge monitoring, particularly when real examples such as prestressing wire breakage are limited.
\end{abstract}


\end{frontmatter}

\nolinenumbers

\noindent\textbf{Impact Statement.}
Bridges are critical infrastructure systems, and rare damage-related acoustic events, such as prestressing wire breakage, are difficult to capture in sufficient quantity under real monitoring conditions. This scarcity limits the availability of representative data for developing and evaluating data-driven structural health monitoring methods. This study investigates GAN-based synthesis of short-time Fourier transform (STFT) spectrograms as a representation-level strategy for enriching limited acoustic-event datasets. By generating synthetic spectrograms that visually and statistically resemble real event representations, the proposed framework provides a basis for studying rare-event data generation in bridge monitoring scenarios. The results demonstrate improved spectrogram fidelity and distributional similarity compared with selected GAN baselines. However, the generated samples are evaluated only at the representation level, and their effect on downstream SHM tasks, such as event classification, damage detection, or anomaly detection, requires further validation in future work.

\section{Introduction}
\label{sec1}
Bridges are vital for supporting economic growth and enabling efficient transportation. However, their structural integrity is continually challenged by factors such as aging, traffic loads, and environmental conditions \cite{ko_technology_2005}. Over time, these influences can lead to material deterioration, making early damage detection through structural health monitoring (SHM) critical for maintenance planning and ensuring structural safety. One severe form of internal deterioration in post-tensioned prestressed concrete bridges is the corrosion of the strands, which can lead to a progressive wire breakage and ultimately lead to catastrophic collapse of the structure, if not detected \cite{ferro_collapse_2022, zingoni_advances_2019}. These incidents underscore the urgent need for proactive monitoring and timely interventions to maintain infrastructure safety and reliability.

Conventional SHM methods, notably visual inspection, remain mainly manual, expensive, and ineffective for detecting internal or subtle structural damage at early stages. In addition, advanced sensor-based methods often produce complex and noisy data, which complicates manual analysis and limits timely decision making \cite{saleem_instant_2021, webb_analysis_2017}. In tackling these challenges, recent attention has turned toward automated processing techniques utilizing Machine Learning (ML) and Deep Learning (DL), which can efficiently interpret large datasets and directly identify structural anomalies from raw data \cite{farhadi_acoustic_2024, zinno_artificial_2022}. Convolutional Neural Networks (CNNs), for instance, have been widely applied to analyze vibration signals, indicating their efficiency in detecting structural anomalies \cite{zhou_ambient_2025, duran_novel_2025, parziale_explainability_2024}. Similar approaches have been applied to detect damage in mechanical and structural components \cite{wang_bearing_2025, guo_towards_2024, azad_hybrid_2024}.

In recent studies \cite{farhadi_acoustic_2024, farhadi_automated_2024, farhadi_prestressing_2024}, the authors investigated the downstream task of sound event detection for classifying damage events, with particular focus on prestressing wire breakage—a critical yet rare and extremely short-lived occurrence in SHM. A comprehensive benchmarking of several deep learning models, including modified pre-trained models and a customized model (AcousticNet), confirmed the superior performance of Short-time Fourier Transform (STFT)-based representations compared to other time-frequency representations. Nevertheless, these studies also underscored persistent challenges: severe data scarcity and class imbalance, both of which hindered model generalization and robustness. Conventional augmentation strategies such as MixUp were explored as partial remedies, but their inherent limitations became clear. In particular, MixUp generates synthetic samples by linear interpolation, often producing unrealistic or ambiguous signals, especially for transient anomalies of very short duration. Consequently, such techniques proved insufficient to address the constraints posed by small, imbalanced, and noisy SHM datasets. This gap motivated the present work: the development of a generative framework for synthesizing realistic STFT representations of underrepresented event classes.

To address these challenges, the present study proposes a customized generative model designed to end-to-end synthesize realistic single-channel STFT spectrograms of bridge monitoring events. The model is based on the Wasserstein GAN with Gradient Penalty (WGAN-GP) architecture to ensure stable training and high-quality generation. Training is performed in an event-separated but unsupervised manner, meaning that although each GAN is trained on a specific event type (e.g., wire breakage or traffic), no class labels or downstream classifiers are involved. This setup enables the model to learn the spectral structure of each event class independently, enriching the dataset for rare classes without relying on manual annotation.

Unlike previous approaches that use three-channel (RGB) spectrograms, this study employs single-channel STFT inputs, simplifying the model and preserving the native structure of vibration data. The main contributions are as follows: (1) Acquisition of a rare, high-quality dataset from controlled wire breakage experiments on a real post-tensioned concrete bridge in Italy slated for demolition, enabling safe, repeatable testing under realistic structural conditions; (2) Development of STFTSynth, a WGAN-GP-based generative model tailored to synthesize realistic single-channel STFT spectrograms of various events; (3) Comprehensive benchmarking of STFTSynth against baseline models (DCGAN, standard WGAN-GP, and LSGAN) using both visual inspection and quantitative metrics. This work demonstrates the potential of GAN-based data generation for representation-level synthesis of rare SHM acoustic-event spectrograms. The contribution is focused on synthetic STFT spectrogram generation and representation-level benchmarking, while the effect of the generated samples on downstream SHM tasks remains a direction for future work.

\section{Background}
Despite the significant potential of deep learning (DL) methods in SHM, several challenges limit their generalization and deployment in real-world scenarios. A major issue is data scarcity, as rare but critical events such as prestressing wire breakage are extremely difficult and costly to capture in sufficient quantity for robust training \cite{farhadi_automated_2024, chen_shifting_2024}. Furthermore, class imbalance, due to the dominance of normal structural states, can bias model performance and reduce sensitivity to damage \cite{gao_balanced_2021}. Real-world datasets are often affected by environmental noise, such as traffic or construction vibrations, which complicate anomaly detection and diminish signal clarity \cite{xie_global_2022, tran_data_2021, bull_probabilistic_2021}. Although modern SHM methods increasingly support real-time and cloud-based monitoring, their underlying models are typically trained offline and require large, labeled datasets. These conditions make supervised DL difficult to scale for critical but infrequent events. To overcome these limitations, generative modeling has appeared as a promising strategy for synthesizing additional representations of underrepresented events, such as wire breakage. However, the practical value of such synthetic samples for SHM deployment requires downstream validation.

Generative Adversarial Networks (GANs) \cite{goodfellow_generative_2014} have been widely adopted for synthetic data generation in audio and vibration domains. Early waveform-based approaches (e.g., WaveGAN \cite{donahue_adversarial_2018}) demonstrated the feasibility of adversarial learning for temporal signals but often suffered from training instability and limited spectral coherence, particularly for complex structural monitoring data. Consequently, later studies shifted toward spectrogram-based GANs, where time–frequency representations such as STFT enable more stable training and improved representation of spectral content (e.g., SpecGAN \cite{donahue_adversarial_2018}, GANSynth \cite{engel_gansynth_2019}, MelGAN \cite{kumar_melgan_2019}, EVA-GAN \cite{liao_eva-gan_2024}). These works collectively support the use of STFT-domain representations as an effective and practical basis for adversarial audio generation.

Recent work has begun exploring the potential of GANs in SHM applications, particularly for generating structural monitoring signals. Luleci et al. \cite{luleci_generative_2023} proposed a one-dimensional Wasserstein Deep Convolutional GAN with Gradient Penalty (1D WDCGAN-GP) to generate synthetic acceleration time-series for damage detection. Their study demonstrated that integrating synthetic samples with real datasets improved the performance of downstream classifiers, particularly in addressing data imbalance. These efforts confirmed the viability of GAN-based augmentation for SHM but remained limited to waveform-level generation. Similarly, Tran et al. \cite{tran_data_2021} explored the use of GANs to generate vibration signals under varying operational conditions, highlighting GANs’ potential to model complex data distributions in noisy SHM setting. Beyond waveform data, GANs have also been employed for structural defect simulation. Li et al. \cite{li_data_2024} applied CycleGAN to generate crack images on bridge surfaces, aiding in image-based defect classification. In related work, Le et al. \cite{le_learning-based_2020} used WGANs to generate surface defect patterns on metallic structures. Although these image-based approaches are promising, they do not address the spectral and temporal complexity inherent in time-series based (vibration-based) SHM. Besides, in parallel to data-driven approaches, physics-informed GAN have indicated potential for sound and demonstrating improved interpretability and fidelity \cite{ren_example-guided_2013, su_physics-driven_2023}.  These studies confirm the application of GANs for structural data augmentation, however, very few works have investigated GAN-based generation of time-frequency representations, particularly single-channel STFT spectrograms, in the context of SHM. This gap is significant, as spectrograms provide a compact, interpretable, and structure-preserving way to represent acoustic signals, especially for rare and short-lived events like prestressing wire breakage. The current study addresses this gap by introducing a GAN-based framework specifically designed to synthesize realistic STFT spectrograms of structural events.

Moreover, several studies have explored generative modeling in SHM context, primarily with the aim of mitigating data scarcity and class imbalance. Existing SHM-focused works have mainly employed GAN-based approaches for waveform-level synthesis of vibration or acceleration signals, including labeled data augmentation and signal reconstruction tasks \cite{hou_deep_2022, luleci_generative_2022, mousavi_transformer-based_2025}. Related efforts have also explored GAN-based augmentation of acoustic emission signals for damage-related event analysis \cite{fu_concrete_2024}. Although these studies demonstrate the potential of generative models in SHM, they mainly operate in the time-domain and do not address the generation of time–frequency representations. In contrast, the present study focuses on the generation of single-channel STFT spectrograms for rare acoustic events, addressing a gap in the current SHM literature where time–frequency–based generative augmentation remains largely unexplored.

\section{Methods}
\label{sec3}
\subsection{Data Preparation}
\label{section 5.1}
\begin{figure*}[!ht]\captionsetup[subfigure]{font=small}
    \centering
    \begin{subfigure}{0.48\textwidth}    
        {\includegraphics[width=\linewidth]{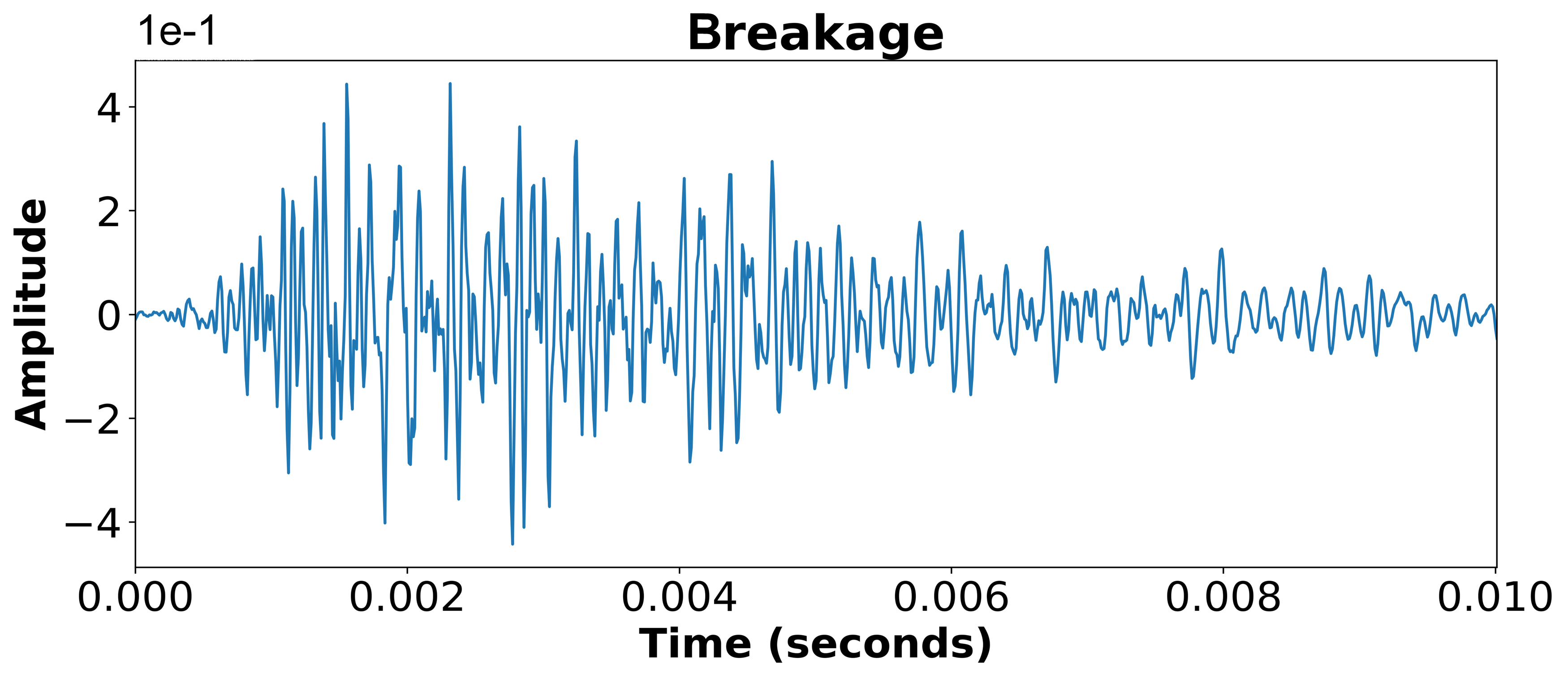}}
        \caption*{(a)}
    \end{subfigure}  
    \hfill
    \begin{subfigure}{0.48\textwidth}
        {\includegraphics[width=\linewidth]{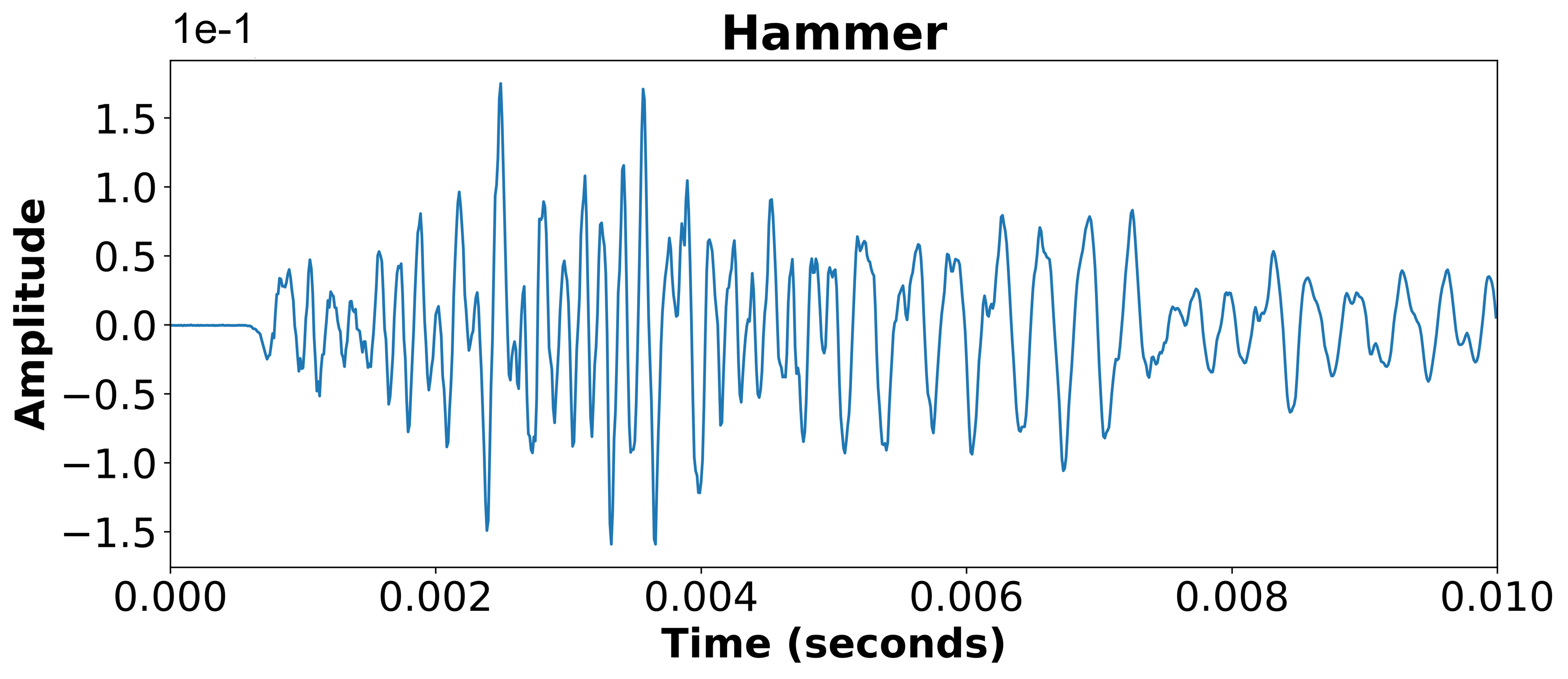}} 
        \caption*{(b)}
    \end{subfigure}
    \hfill 
    
    \begin{subfigure}{0.48\textwidth}
        {\includegraphics[width=\linewidth]{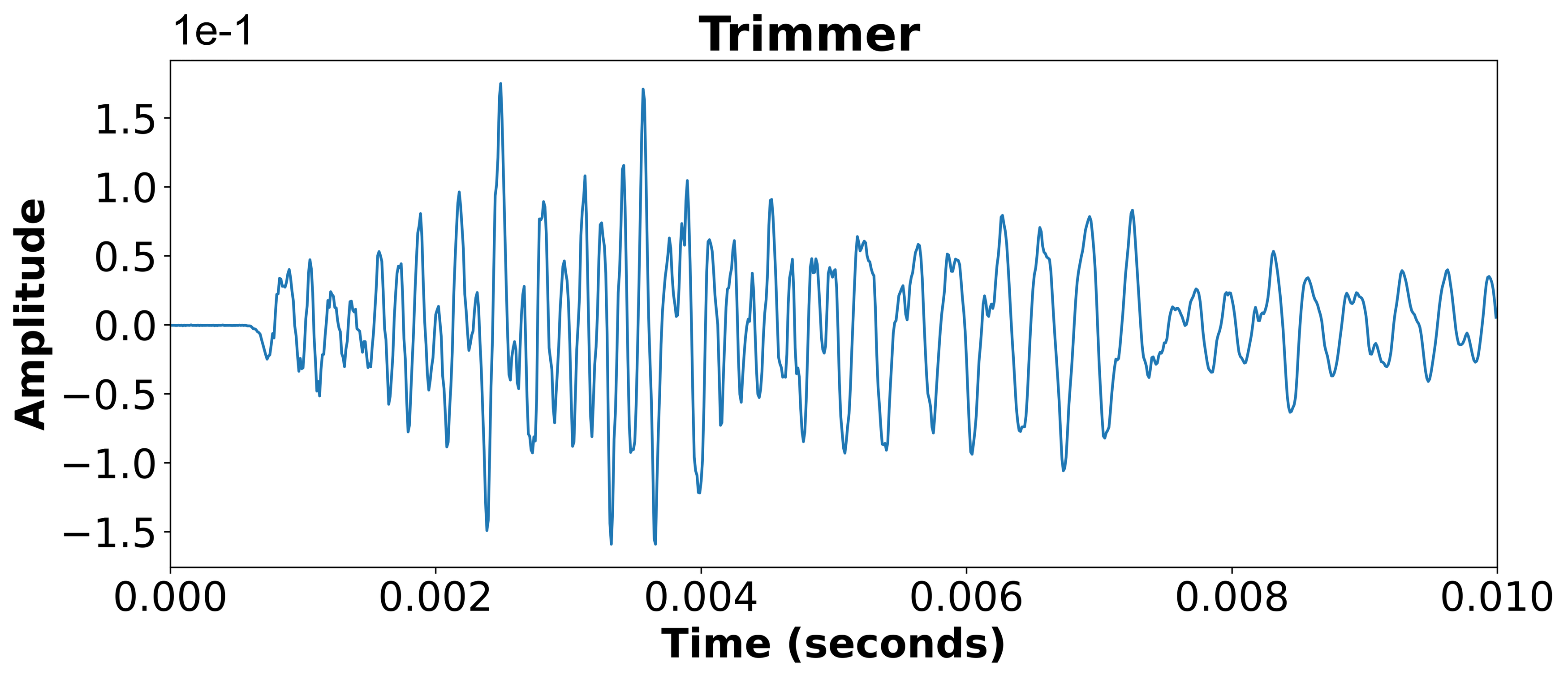}} 
        \caption*{(c)}
    \end{subfigure}
    \hfill 
    \begin{subfigure}{0.48\textwidth}
        {\includegraphics[width=\linewidth]{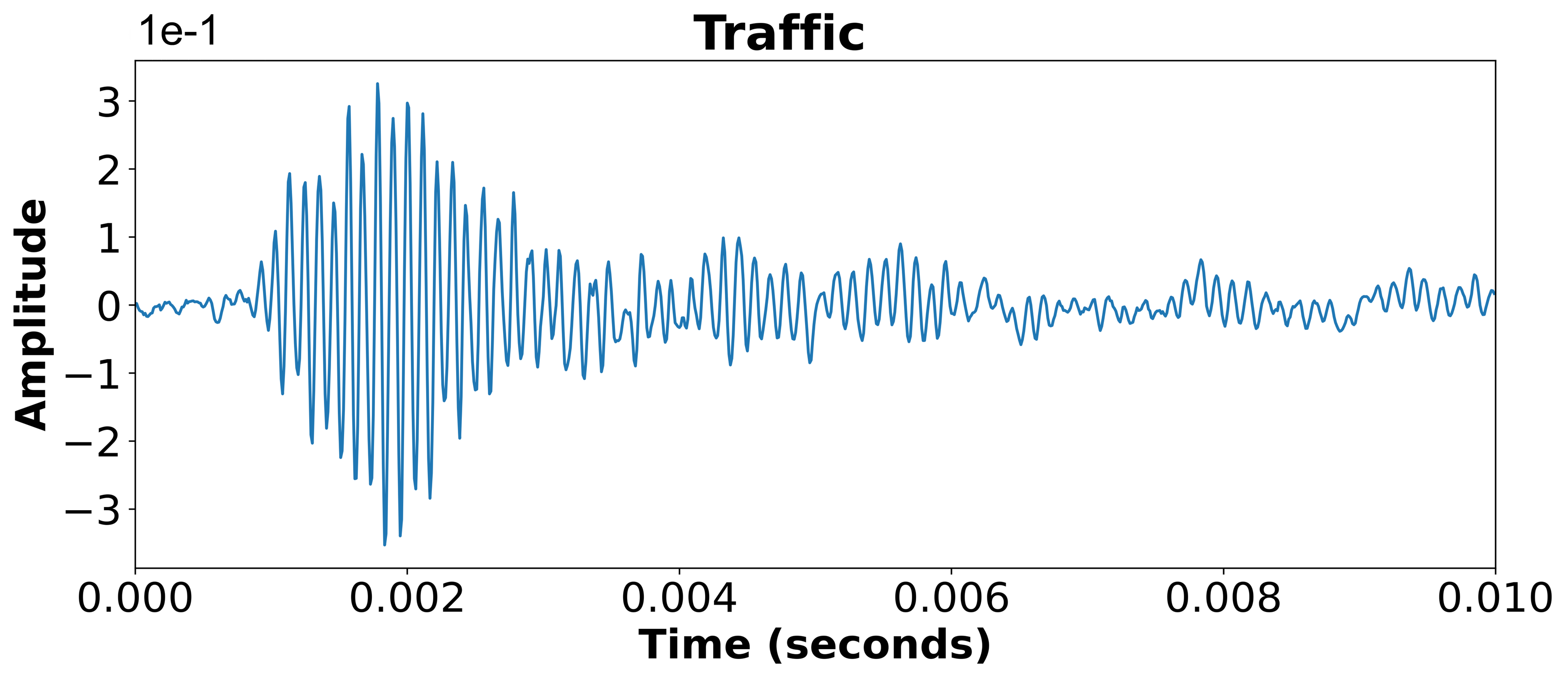}} 
        \caption*{(d)}
    \end{subfigure}
    \hfill 
    \caption{Time-domain representation of the recorded acoustic events}
 \label{fig:time_freq_gan}
\end{figure*}

Starting from the continuous recordings, single acoustic events with a duration of 0.0104 s were extracted. Considering the acquisition rate of 96 kHz, each signal consists of 1000 samples. The length of the signals was chosen based on experimental observations and numerical modelling, which indicate that the duration of a breakage event is approximately 0.01 s. As discussed earlier, high-quality labeled data are critical for enhancing model robustness. In this study, the dataset comprises four signal classes: wire breakage (202 signals), hammering (264 signals), electric trimmer (459 signals), and traffic noise (415 signals). The distribution highlights a class imbalance, with critical events such as wire breakage being underrepresented. This imbalance can adversely affect downstream classification model performance by biasing predictions toward the more frequent classes.
Figure \ref{fig:time_freq_gan} provides an example of the representations in the time domain of each signal class, highlighting the distinctive patterns captured during the acquisition process. To ensure uniformity and meaningful comparison, all signals were normalized before further processing. This preprocessing step reduces the bias from varying amplitude ranges and can simplify model training. The next step is to extract features that can capture necessary information for model training, as discussed in the next section.
%
\subsection{Feature Extraction}
\label{section 5.2}
\begin{figure*}[!ht]
    \centering
    \includegraphics[width=\textwidth]{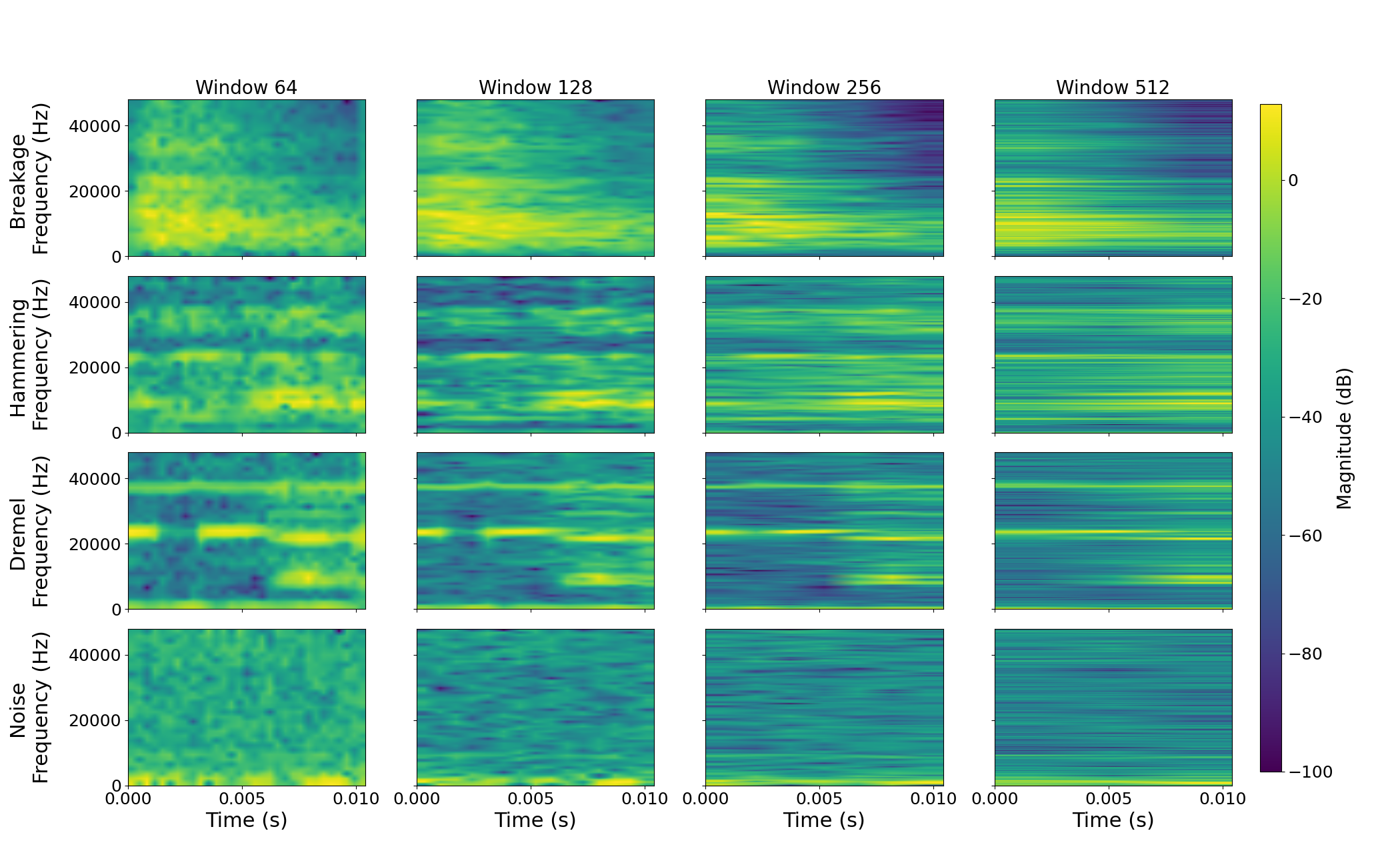}
    \caption{STFT spectrograms of events across four window sizes, illustrating the impact of resolution on signal representation and feature extraction}
    \label{fig:window_sizes_events}
\end{figure*}

Short-Time Fourier Transform (STFT) was employed as the primary feature extraction method for in-situ acoustic signals, transforming them from the time domain into the time–frequency domain to capture both spectral and temporal characteristics. This choice aligns with prior work by the authors and supports dynamic signal representation for GAN-based data generation. All STFT computations employed a Hann window function with 50\% overlap. This choice was supported by preliminary tests, which indicated that Hann windows produced stable and informative spectrograms for wire breakage signals, while a 50\% overlap maintained sufficient temporal resolution without introducing excessive redundancy \\
Window size is a key STFT parameter, directly affecting the time–frequency resolution trade-off (\textbf{Gabor limit}). Figure~\ref{fig:window_sizes_events} compares spectrograms for four window sizes: small windows (e.g., 64 samples) yield high temporal resolution but poor frequency resolution, while large windows (e.g., 512 samples) achieve the opposite. To balance these effects, window sizes of 128 and 256 were selected, offering both spectral clarity and temporal precision. The resulting spectrogram dimensions, expressed as (frequency bins, time windows), were $(65,15)$ and $(129,7)$, respectively. All STFT outputs were treated as single-channel images during training to reduce computational complexity.

%

\subsection{\textsc{Sequence Modeling: Gated Recurrent Unit}}
Sequential modeling was initially developed to address the limitations inherent in Convolutional Neural Networks (CNNs), particularly their inability to capture long-term dependencies in sequential data. Early models include Jordan Networks \cite{jordan_serial_1986} and Elman Networks \cite{elman_finding_1990}, both foundational forms of Recurrent Neural Networks (RNNs) designed to process sequential information. Among the more advanced architectures, Gated Recurrent Units (GRUs) \cite{cho_properties_2014} were introduced as a computationally efficient alternative to Long Short-Term Memory (LSTM) networks. GRUs incorporate gating mechanisms that effectively preserve long-range dependencies and mitigate the vanishing gradient problem. By combining the roles of LSTM’s input and forget gates into a single update gate, GRUs reduce model complexity while maintaining comparable performance for various sequence modeling tasks (Figure \ref{fig:bigru}).

\begin{figure}[!ht]
    \centering
    \resizebox{0.75\columnwidth}{!}{
    \begin{tikzpicture}
    font=\sf \scriptsize,
        \tikzstyle{cell}=[draw, very thick, shape=rectangle, minimum size=1.15cm, rounded corners=5mm],
        \tikzstyle{node}=[draw, thick, shape=circle, minimum size=0.8cm],
        \tikzstyle{operator}=[draw, thick, shape=circle, inner sep=0.3mm],
        \tikzstyle{gate}=[draw, thick, shape=rectangle, minimum size=1cm, rounded corners=1mm],

        \tikzstyle{line}=[rounded corners=.15cm, thick],
        
        \node [cell, minimum height=4cm, minimum width=6cm] {};

        \node [node](ht1) at (-4, 1.5){$h_{t-1}$};
        \node [node](xt) at (-2.5, -3){$x_{t}$};

        \node [node](ht) at (4, 1.5){$h_{t}$};
        \node [node](y) at (2.5, 3){$y$};

        \node [](r_t) at (-1.2, 0.2){$r_{t}$};
        \node [](z_t) at (0, 0.2){$z_{t}$};
        \node [](h_t) at (2, -0.4){$\tilde{h}_{t}$};

        \node [operator](ew1) at (-2.25, 0){$\times$};
        \node [operator](ew2) at (0.25, 1.5){$\times$};
        \node [operator](ew3) at (1.75, 0){$\times$};
        \node [operator](add) at (1.75, 1.5){+};
        \node [operator](negative) at (0.25, 0.75){1-};

        \node [gate, minimum height=0.5cm, minimum width=0.8cm](sigma1) at (-1.25, -1){$\sigma$};
        \node [gate, minimum height=0.5cm, minimum width=0.8cm](sigma2) at (0.25, -1){$\sigma$};
        \node [gate, minimum height=0.5cm, minimum width=0.8cm](tanh) at (1.75, -1){$\tanh$};

        \draw[line] (xt) |- (0, -1.75);
        \draw [line] (ht1) |- (0.25, -1.5);
        \draw [line] (ht1 -| sigma1)++(-1,-3.25) -| (sigma1); 
        \draw [line] (ht1 -| sigma2)++(-1,-3.25) -| (sigma2);
        \draw [line] (xt -| tanh)++(-1.75, 1.25) -| (tanh); 

        \draw[line] (ew1) |- (-1.85, -1.75); 
        
        \draw[->, line] (ht1) -- (ew2) -- (add) -- (ht);
        \draw[->, line] (add) -| (y);

        \draw [->, line] (sigma1) |- (ew1);
        \draw [line] (ew1) |- (0, 1.5);

        \draw [->, line] (sigma2) -- (negative) -- (ew2);
        \draw [line] (sigma2) |- (1.55, 0);
        
        \draw [->, line] (tanh) -- (ew3) -- (add);

    \end{tikzpicture}
    }
    \caption{Architecture of a Gated Recurrent Unit (GRU)}
    \label{fig:bigru}
\end{figure}

At each time step $t$, the hidden state $h(t)$ is updated using a linear interpolation between the previous hidden state $h(t-1)$ and the candidate activation $\tilde{h}(t)$, weighted by the update gate $z(t)$. The updated equation is given as \cite{cho_learning_2014, chung_empirical_2014}:  
\begin{equation}
    \label{eq:gru_hidden_state}
    h(t) = (1 - z(t)) \otimes h(t-1) + z(t) \otimes \tilde{h}(t)
\end{equation}
where $\otimes$ represents element-wise multiplication, and $z(t)$, known as the update gate, determines the extent to which the hidden state should be updated with new information. The update gate is computed as:
\begin{equation}
    \label{eq:reset_gate}
    z(t) = \sigma(w_{z} \cdot x(t) + w_{h} \cdot h(t-1) + b(z))
\end{equation}
where $w_z$ and $w_h$ are the weight matrices for the input and hidden state, respectively, and $b_z$ is the bias vector. The update gate in GRU simultaneously performs the role of the input and output gates in LSTM. The sigmoid function, $\sigma$, constrains $z(t)$ to values between 0 and 1, determining the proportion of $h(t-1)$ that should be preserved versus the amount of $\tilde{h}(t)$ (candidate activation) to incorporate. This update gate effectively regulates the flow of information from the previous hidden state to the current one. The candidate activation $\tilde{h}(t)$ is computed similarly to the hidden state in a conventional RNN, with an additional modulation by the reset gate $r(t)$ as follows:
\begin{equation}
    \label{eq:candidate_activation}
    \tilde{h}(t) = \tanh(w_{x} \cdot x(t) + w_{\tilde{h}} \cdot (r(t) \otimes h(t-1)) + b_{\tilde{h}})
\end{equation}
here, $w_{x}$ and $w_{\tilde{h}}$ are weight matrices for the input and the reset-modulated hidden state, respectively. $r(t)$ is known as the \textit{reset gate}, determines how much of the previous hidden state $h(t-1)$ can be retained when calculating the candidate activation. The reset gate is computed as:
\begin{equation}
    r(t) = \sigma(w_{x} \cdot x(t) + w_{h} \cdot h(t-1) + b_{r})
\end{equation}
where $w_{x}$ and $w_{h}$ are the weights for the input and hidden state in the reset gate, and $b_r$ is the corresponding bias vector. 

\subsection{\textsc{Generative Adversarial Networks}}
Generative Adversarial Networks (GANs), first introduced by Goodfellow et al. in 2014 \cite{goodfellow_generative_2014}, are generative models that learn to produce data resembling a target distribution through an adversarial process. A GAN consist of two neural networks: the generator, which maps random latent vectors ($z \sim p_z(z)$) to synthetic samples ($\tilde{x}$) and the discriminator (or critic) which distinguish real samples ($x \sim p_{data}$) from the generated ones (Fig. \ref{fig:gan_workflow}). The two networks are trained in a minimax game \cite{gui_review_2023, xia_gan-based_2022}:
\begin{equation}
    \label{eq:obj_func}
    \min_{G} \max_{D} \mathbb{E}_{x \sim p_{\text{data}}}[\log D(x)] + \mathbb{E}_{z \sim p_z(z)}[\log(1 - D(G(z)))]
\end{equation}
where $D(x) \in (0,1)$ is the discriminator's estimated probability that $x$ is real, $p_{\text{data}}$ is the distribution of the real dataset, and $p_z(z)$ is the prior distribution over latent vectors (often Gaussian or uniform). The generator $G$ aims to minimize this objective by producing realistic samples that maximize $D(G(z))$, while the discriminator $D$ aims to maximize it by correctly distinguishing real from generated data.

\begin{figure*}
    \centering
    \resizebox{0.85\columnwidth}{!}{
    \begin{tikzpicture}[node distance=2cm and 2cm]

    \tikzstyle{data} = [rectangle, minimum width=3cm, minimum height=3cm, text centered, draw=black, thick, rounded corners, fill=cyan!30]
    \tikzstyle{noise} = [rectangle, minimum width=2cm, minimum height=1cm, rounded corners, text centered, draw=black]
    \tikzstyle{generator} = [rectangle, minimum width=3cm, minimum height=3cm, text centered, draw=black, thick, rounded corners, fill=Mulberry!30]
    \tikzstyle{discriminator} = [rectangle, minimum width=3cm, minimum height=3cm, text centered, draw=black, thick, rounded corners, fill=Periwinkle!30]
    \tikzstyle{dashedrect} = [draw=black, dashed, thick, inner sep=0.3cm, rectangle, rounded corners]
    \tikzstyle{arrow} = [thick, ->, >=stealth]
    \tikzstyle{backprop} = [thick, ->, >=stealth, red!70, line width=0.3mm]

    \node (generator) [generator] {Generator};
    \node (random_noise) [noise, left=of generator, rotate=0, yshift=0cm, xshift=1.5cm] {\shortstack{Random\\Noise}};
    \node (data) [data, below=of generator] {Training Data};
    \node (fake_images) [right=2cm of generator, rotate=0, xshift=-1.2cm] {\shortstack{Fake \\ Images}};
    \node (real_images) [right=2cm of data, rotate=0, xshift=-1.2cm] {\shortstack{Real \\ Images}};
    \node (discriminator) [discriminator, right=of data, xshift=1.15cm, yshift=2.5cm] {Discriminator};
    \node (gen_loss) [right=1.5cm of discriminator, rotate=0, yshift=2.5cm, xshift=-0.25cm] {\shortstack{Generator \\ Loss}};
    \node (disc_loss) [right=1.5cm of discriminator, rotate=0, yshift=-2.5cm, xshift=-0.58cm] {\shortstack{Discriminator \\ Loss}};

    \node (backprop) [right=of generator, xshift=3cm, yshift=0.5cm] {backprop};
    \node (backprop) [right=of data, xshift=3cm, yshift=-0.5cm] {backprop};
    \node (backprop) [right=of data, xshift=-1.5cm, yshift=7.3cm] {backprop};
    
    \node[dashedrect, fit={(fake_images)}] (fake_dashed) {};
    \node[dashedrect, fit={(real_images)}] (real_dashed) {};
    \node[dashedrect, fit={(gen_loss)}] (gen_loss) {};
    \node[dashedrect, fit={(disc_loss)}] (disc_loss) {};

    \draw [arrow] (random_noise.east) |- (generator.west);
    \draw [arrow] (generator.east) |- (fake_dashed.west);
    \draw [arrow] (data.east) |- (real_dashed.west);
    \draw [arrow] (fake_dashed.south) |- (discriminator.west); 
    \draw [arrow] (real_dashed.north) |- (discriminator.west); 
    \draw [arrow] (discriminator.east) -| (gen_loss.south); 
    \draw [arrow] (discriminator.east) -| (disc_loss.north);

    \draw [backprop] (disc_loss.west) -| (discriminator.south);
    \draw [backprop] (gen_loss.west) -| (discriminator.north);
    \draw [backprop] 
    ([xshift=-0.1cm] discriminator.north) |- 
    ([xshift=3cm] fake_dashed.north) |- 
    ([yshift=0.5cm] generator.north) -| 
    (generator.north);

    \end{tikzpicture}
    }
    \caption{Overview of GAN workflow.}
    \label{fig:gan_workflow}
\end{figure*}

In practice, vanilla GANs often suffer from unstable training and mode collapse. Variants such as Deep Convolutional GAN (DCGAN) \cite{radford_unsupervised_2016} introduced convolutional architectures, Least Squares GAN (LSGAN) \cite{mao_least_2017} replaced the cross-entropy loss with a least-squares loss to improve gradient quality, and Wasserstein GAN (WGAN) \cite{arjovsky_wasserstein_2017} employed the Wasserstein distance for more stable optimization. The WGAN with Gradient Penalty (WGAN-GP) \cite{gulrajani_improved_2017} further improved stability by enforcing a soft 1-Lipschitz constraint via a gradient penalty term:
\begin{equation}
    \label{eq:wgan}
    \mathcal{L}_D = \mathbb{E}_{\tilde{x} \sim \mathbb{P}_g} [D(\tilde{x})] - \mathbb{E}_{x \sim \mathbb{P}_r} [D(x)] + \lambda \mathbb{E}_{\hat{x} \sim \mathbb{P}_{\hat{x}}} \left[(\|\nabla_{\hat{x}} D(\hat{x})\|_2 - 1)^2\right]
\end{equation}
where $\mathbb{P}_r$ and $\mathbb{P}_g$ are the real and generated distributions, $\hat{x}$ is an interpolation between real and fake samples, and $\lambda$ is the penalty coefficient. WGAN-GP produces smoother gradients and reduces mode collapse \cite{bishop_deep_2024}, making it suitable for high-dimensional data generation.

\subsection{\textsc{Proposed GAN Framework}}
Convolutional GANs such as DCGAN and LSGAN are effective at modeling local spatial structure in spectrograms. However, without explicit sequential modeling, they may not fully capture long-range temporal dependencies across frames, particularly when events span multiple STFT windows. Sequence-aware components can complement convolutional layers to improve temporal consistency. To address this limitation, this research proposed STFTSynth architecture which integrates convolutional residual blocks for spatial feature extraction with bidirectional Gated Recurrent Units (BiGRUs) to explicitly model temporal dependencies in both forward and backward directions. The residual blocks enhance gradient flow and enable multi-scale feature learning. On the other hand, BiGRUs capture the evolution of spectral patterns over time, ensuring coherent temporal transitions between adjacent STFT frames.\\
STFTSynth is implemented within the WGAN-GP framework to utilize its stable training dynamics and improved gradient behavior. The generator combines dense residual connections with BiGRUs to produce spectrograms that contain realistic local spectral texture and consistent temporal structure. The discriminator employs spectral normalization and multi-scale feature extraction to improve robustness against overfitting and mode collapse. This hybrid design enables STFTSynth to generate high-fidelity, temporally consistent STFT spectrograms for rare SHM events, such as prestressing wire breakage, that are difficult to capture in sufficient quantity for supervised training.

The overall workflow of the proposed GAN-based STFT data generation is illustrated in Fig.~\ref{fig:stftsynth_workflow}. It outlines the progression from raw event collection to feature extraction, GAN-based spectrogram synthesis, and evaluation using quantitative and qualitative metrics. Implementation details, training procedures, and hyperparameter settings of the proposed architecture are provided in Section \ref{section 5.3}

\begin{figure}[H]
    \centering

    \tikzstyle{block} = [rectangle, draw, fill=white, 
                         text centered, rounded corners, minimum height=3em, minimum width=8em]

    \begin{tikzpicture}[scale=0.85, every node/.style={align=center}, node distance=2cm]
        \node [block] (DCol) {\large{Data Collection} \\ \footnotesize{Wire Breakage, Hammer, Trimmer, Traffic}};

        \node [block, below of = DCol] (STFT) {\large{Feature Extraction} \\ \footnotesize{Short-time Fourier Transform (STFT)}};
        
        \node [block, below left=1cm and -2.75cm of STFT] (DCGAN) {DCGAN};
        \node [block, right of = DCGAN, node distance = 3.5cm] (WGAN) {WGAN-GP};
        \node [block, below of = DCGAN, node distance = 1.5cm] (LSGAN) {LSGAN};
        \node [block, below of = WGAN, node distance = 1.5cm] (Synth) {STFTSynth};

        \node[draw=black, thick, dotted, rounded corners, inner xsep=1em, inner ysep=1em, 
              fit=(DCGAN)(WGAN)(LSGAN)(Synth)] (boxGANs) {};
        \node[fill=white] at (boxGANs.south) {GAN Models};

        \node [block, below=1cm of boxGANs] (Gen) {\large{Synthetic STFT Generation} \\ \footnotesize{Resolution Depends on STFT Window Size}};

        \node [block, below of = Gen, node distance = 2cm] (Eval) {\large{Evaluation} \\ \footnotesize{FID, SSIM, PSNR, Visual Inspection}};

        \draw[->, thick] ([yshift=-0.2em]DCol.south) -- (STFT);
        \draw[->, thick] ([yshift=-0.2em]STFT.south) -- (boxGANs);
        \draw[->, thick] ([yshift=-1em]boxGANs.south) -- (Gen);
        \draw[->, thick] ([yshift=-0.2em]Gen.south) -- (Eval);

    \end{tikzpicture}

    \caption{Overview of the proposed GAN-based data augmentation workflow using STFT spectrograms.}
    \label{fig:stftsynth_workflow}
\end{figure}

\subsection{Model Developments and Training Procedure}
\label{section 5.3}
This section presents the development, implementation, and training procedures for four GAN models: DCGAN, WGAN-GP, LSGAN, and the proposed STFTSynth. The baseline models were selected due to their widespread adoption and established performance in image generation tasks, providing a reliable foundation for comparison. However, architectures such as DCGAN are known to suffer from issues including mode collapse and training instability, particularly when applied to small and imbalanced datasets. To address these limitations, WGAN-GP and LSGAN introduce alternative loss formulations that improve training stability and enhance sample diversity. Despite these improvements, the baseline models still exhibit limitations in capturing long-range temporal dependencies present in STFT spectrograms. This limitation motivates the development of STFTSynth, a task-specific architecture designed to improve both temporal consistency and spectral realism. Highly complex architectures such as StyleGAN were not considered due to their significant computational requirements and dependence on large-scale, diverse datasets, which are not available in this study. Additionally, hyperparameter tuning was performed to further mitigate issues related to mode collapse and overfitting. All models were trained using a dataset collected from in-situ bridge monitoring.

The implementation environment and computational setup are described as follows. Training experiments were conducted on a workstation equipped with an NVIDIA RTX A6000 GPU (48 GB memory). GPU acceleration enabled the customized models to complete each training epoch in approximately 15 seconds on average, depending on model complexity and batch size. In contrast, the selected baseline models required less than 2 seconds per epoch. All models were implemented using Keras with a TensorFlow 2.10 backend, within a Python 3.9 environment and its associated scientific computing libraries. The workstation operated on Windows Server 2019, with GPU performance optimized using CUDA Toolkit 11.2 and cuDNN 8.0.

\paragraph{\textbf{DCGAN}} 
The DCGAN implementation followed Radford et al. \cite{radford_unsupervised_2016}, adapted for single-channel STFT spectrograms (size = H × W). The generator starts from a 100-dimensional latent vector, followed by four transposed convolutional layers (kernel = 4, stride = 2, padding = 1) with batch normalization and Leaky ReLU activations, producing the target resolution. The discriminator mirrors this structure in reverse using strided 2D convolutions with spectral normalization and Leaky ReLU activations, ending with a sigmoid output for real/fake classification. All optimizer settings, initialization schemes, and other training hyperparameters are summarized in Table \ref{tab:gan_param_training}.

\begin{table*}[!ht]
    \centering
    \caption{Summary of common training hyperparameters and configurations for GAN variants}
    \label{tab:gan_param_training}
    \scriptsize
    \renewcommand{\arraystretch}{1.5} 
    \begin{tabularx}{0.9\textwidth}{@{}l X X@{}}
    \hline
    \textbf{Parameter} & \textbf{Optimal Value} & \textbf{Search Space} \\ \hline 

    \multicolumn{3}{l}{\textbf{General Training Parameters}} \\ \hline
    Learning Rate (Generator) & 2e-5 & 5e-7 to 1e-3 \\
    Learning Rate (Discriminator) & 2e-6 & 5e-7 to 1e-4 \\
    Learning Rate Adjustments & Factor: 0.15, Patience: 5 & Factor: 0.1 to 0.5, Patience: 3 to 10\\ 
    Batch Size & 16 & 16 to 64 \\ 
    Epochs & Max 2000 & 1000 to 5000 \\ 
    Optimizer & Nadam (Adam with Nesterov momentum) & Adam, Nadam, RMSprop \\  
    Weight Initialization & Xavier & Xavier and He \\ 
    Latent Space Dimension \( Z \) & 100 & 50 to 500 \\  
    Regularization & Batch Normalization, Dropout rate: 0.2 & Dropout: 0.1 to 0.4 \\  

    \hline
    \multicolumn{3}{l}{\textbf{GAN-Specific Parameters}} \\ \hline
    Loss Function (DCGAN) & Binary Cross-Entropy (BCE) & BCE \\ 
    Loss Function (LSGAN) & Least Squares Loss & Least Squares, Huber Loss \\ 
    Loss Function (WGAN-GP) & Wasserstein Loss with Gradient Penalty & Wasserstein Loss \\ 
    Gradient Penalty \( \lambda_{GP} \) (WGAN-GP) & 12 & 1 to 20 \\ 
    Gradient Clipping (if applicable) & Not used & 0.01 to 0.1 \\ 
    Activation Function (Generator) & Leaky ReLU (all layers except last), Tanh & ReLU, Leaky ReLU, ELU, Tanh \\ 
    Activation Function (Discriminator) & Leaky ReLU (all layers except last), Sigmoid & ReLU, Leaky ReLU, ELU, Sigmoid \\  
    Generator Layers & 4 Transposed Convolutions & 3 to 8 \\  
    Discriminator Layers & 5 Convolutional Layers & 4 to 8 \\  
    Kernel Size (Conv Layers) & 5 × 5 & 3 × 3 to 5 × 5 \\  
    Stride (Conv Layers) & 2 & 1 to 2 \\  

    \hline
    \end{tabularx}
\end{table*}
%

\paragraph{\textbf{WGAN-GP}} 
The WGAN-GP model retained the same convolutional generator–discriminator architecture as DCGAN for a fair comparison but replaced the BCE loss with the Wasserstein loss and added a gradient penalty ($\lambda = 12$) to enforce the 1-Lipschitz constraint. The discriminator (critic) outputs real-valued scores rather than probabilities, and spectral normalization was used to improve stability. Both generator and critic employed Leaky ReLU activations in all hidden layers, with Tanh at the generator output. Training followed the optimizer and regularization settings listed in Table \ref{tab:gan_param_training}.

\paragraph{\textbf{LSGAN}} 
The LSGAN architecture also matched DCGAN in both generator and discriminator structure but replaced the BCE loss with a least-squares loss to improve gradient quality and reduce saturation. The discriminator minimized the mean squared error between real samples and label 1, and fake samples and label 0. Leaky ReLU activations were used in all hidden layers, with Tanh at the generator output. All other training settings followed Table \ref{tab:gan_param_training}.

\begin{figure*}[!ht]
    \centering
    \rotatebox[]{90}{
    \includegraphics[width=0.9\textheight]{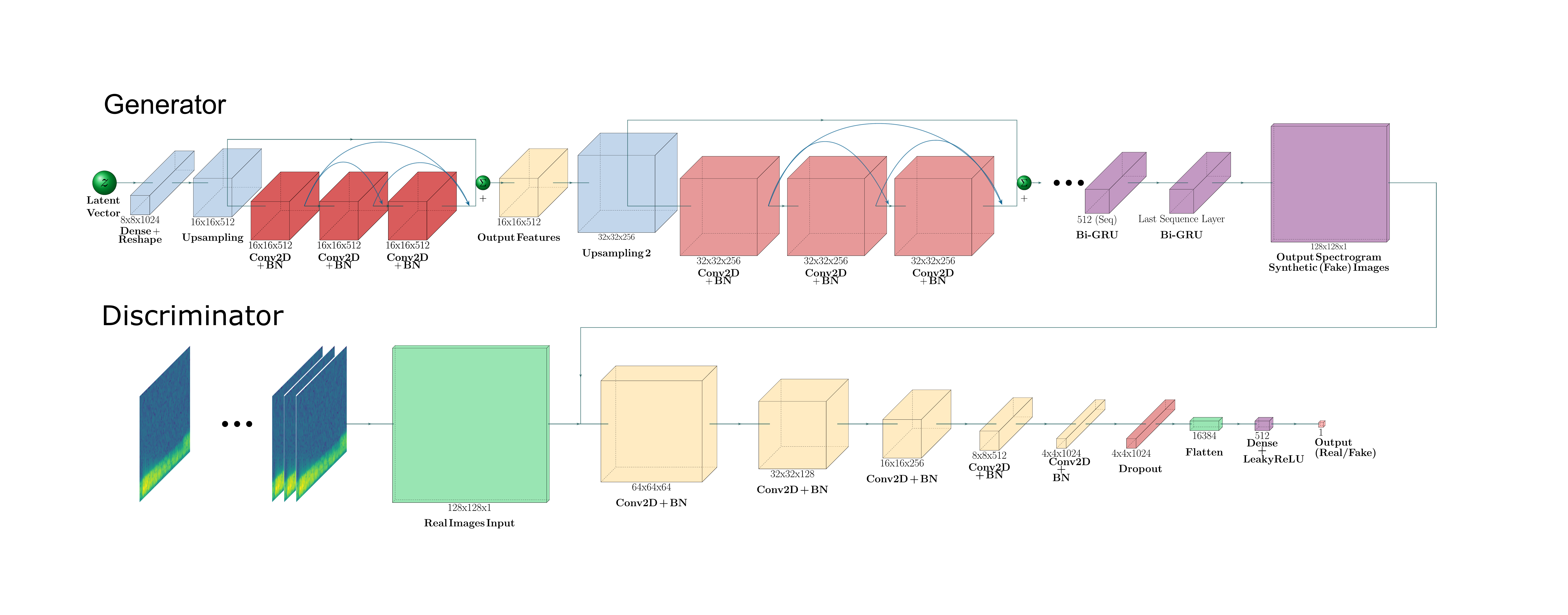}
    }
    \caption{Architecture of the proposed GAN model, STFTSynth}
    \label{fig:stftsynth}
\end{figure*}

\paragraph{\textbf{STFTSynth}} 
STFTSynth extends the WGAN-GP architecture with dense residual blocks (DRBs) and BiGRU layers to address the strong temporal dependencies in STFT spectrograms of structural health monitoring events. DRBs are inserted after upsampling stages in the generator to enhance gradient flow, encourage feature reuse, and integrate multi-scale spectral details. BiGRUs are positioned after intermediate feature maps to model temporal evolution in both forward and backward directions, improving coherence across consecutive STFT frames. The discriminator employs spectral normalization and multi-scale convolutional layers to improve robustness against overfitting and mode collapse. Training follows the WGAN-GP scheme with $\lambda=12$ and matched optimizer settings. A full architectural specification is given in Table~\ref{tab:gan_architectures}, and a schematic is shown in Fig.~\ref{fig:stftsynth}.

\begin{table*}[!ht]
    \centering
    \caption{Comparison of architectural components across GAN variants}
    \label{tab:gan_architectures}
    \scriptsize
    \renewcommand{\arraystretch}{1} 
    \begin{tabularx}{0.9\textwidth}{@{}l X X X X@{}}
    \hline
    \textbf{Component} & \textbf{DCGAN} & \textbf{WGAN-GP} & \textbf{LSGAN} & \textbf{STFTSynth} \\ \hline
    Generator Layers & 4 ConvTranspose2D & 4 ConvTranspose2D & 4 ConvTranspose2D & 4 ConvTranspose2D, 4 DRB and 2 Bi-GRU \\
    Discriminator Layers & 5 Conv2D & 5 Conv2D & 5 Conv2D & 5 Conv2D \\
    Normalization & BatchNorm & BatchNorm & BatchNorm & BatchNorm \\
    Activation (Generator) & Leaky ReLU + Tanh & Leaky ReLU + Tanh & Leaky ReLU + Tanh & Leaky ReLU + Tanh \\
    Activation (Discriminator) & Leaky ReLU & Leaky ReLU & Leaky ReLU & Leaky ReLU \\
    Loss Function & BCE & Wasserstein + GP & Least Squares & Wasserstein + GP \\
    \hline
    \end{tabularx}
\end{table*}
%

The dataset used for training consist of four event classes: wire breakage (202 samples), hammering (264 samples), electric trimmer (459 samples), and traffic noise (415 samples). For each class, a small portion of samples (approximately 15\%) was held out as a validation set and used exclusively for evaluation. Each GAN model was trained under multiple epoch settings (1000–5000) to evaluate training stability and sample quality. Although mode collapse occasionally occurred in early or mid-stages, extended training generally allowed the generators to recover and produce diverse, high-quality spectrograms. Based on comparative evaluations (Table~\ref{tab:gan_param_training}), 2000 epochs were selected as the optimal configuration, offering a balance between stability, fidelity, and computational cost.

\subsection{\textsc{Evaluation Metrics}}

\subsubsection{Structural Similarity Index Measure}
The Structural Similarity Index Measure (SSIM) is a perception-based image quality metric that evaluates the similarity between two images by considering structural information \cite{li_storygan_2018}. In contrast to conventional metrics such as Mean Square Error (MSE), SSIM accounts for the inter-dependencies of neighboring pixels, aligning more closely with human visual perception. SSIM consists of three components, including luminance $I(x,y)$, contrast $C(x,y)$, and structure $S(x,y)$:
\begin{equation}
    \label{eq:ssim}
    \text{SSIM}(x,y) = I(x, y) \cdot C(x, y) \cdot S(x, y)
\end{equation}
In Equation (\ref{eq:ssim}) \cite{zhou_wang_image_2004}:
\begin{equation*}
    I(x,y) = \frac{2 \mu_x \mu_y + c_1}{\mu_x^2 + \mu_y^2 + c_1}
\end{equation*}

\begin{equation*}
    C(x, y) = \frac{2 \sigma_x \sigma_y + c_2}{\sigma_x^2 + \sigma_y^2 + c_2}
\end{equation*}

\begin{equation*}
    S(x, y) = \frac{\sigma_{xy} + c_3}{\sigma_x \sigma_y + c_3}
\end{equation*}

Here, $x$ and $y$ are image patches extracted from real and generated images, respectively. $\mu_x$ and $\mu_y$ are the mean intensity values. $\sigma_x$ and $\sigma_y$ are the standard deviations of the pixel values that represent the contrast. $\sigma_{xy}$ is the covariance between the corresponding pixel values that capture structural similarity. $c_1$, $c_2$, and $c_3$ are small constants to avoid division by zero \cite{zhou_wang_image_2004}. SSIM score ranges between 0 and 1, where a value closer to 1 means more similarity to the real images. 

\subsubsection{Peak Signal-to-Noise Ratio}
The Peak Signal-to-Noise Ratio (PSNR) measures the similarity between two monochrome images to assess the quality of a generated image compared to a real one \cite{smith_scientist_1997}. A higher value of PSNR indicates better image quality. PSNR is computed as follows:

\begin{equation}
    \label{eq:psnr}
    \text{PSNR(R, G)} = 10 \log_{10} (\frac{\text{MAX}^2_I}{\text{MSE}})
\end{equation}
In Equation (\ref{eq:psnr}), \( R \) and \( G \) refer to the real and generated images, respectively, and \( \text{MAX}_I \) is the maximum pixel value of the image and Mean Square Error (\text{MSE}) can be computed as follows:
\begin{equation*}
    \text{MSE} = \frac{1}{mn} \sum _ {i=0}^{m-1} \sum _ {i=0}^{n-1}(R(m, n) - G(m, n)) ^ 2
\end{equation*}

\subsubsection{Fréchet Inception Distance}
The Fréchet Inception Distance (FID) was introduced by Heusel et al. \cite{heusel_gans_2017} to evaluate the performance of GAN models. It measures the difference between real and generated data distributions in a feature space extracted from a pre-trained model, such as Inception-v3. The FID score is computed as follows:

\begin{equation}
    \label{eq:fid}
    \text{FID} = \parallel \mu_{r} - \mu_{g} \parallel ^ 2 + Tr(\Sigma_r + \Sigma_g - 2(\Sigma_r\Sigma_g)^{\frac{1}{2}}) 
\end{equation}
In Equation (\ref{eq:fid}), $\mu_r$, $\Sigma_r$ and $\mu_g$, $\Sigma_g$ are the mean and covariance matrices of real and generated image features, respectively. A lower FID score (typically less than 10 for high-quality images) indicates better image quality and diversity.

\section{Data Acquisition and Problem Setting}
\label{sec4}
Developing an effective automated SHM system requires a robust data acquisition strategy to accurately capture vibration signals under real-world conditions. High-quality, well-labeled datasets are particularly critical for tasks such as synthetic data generation and structural event classification. This section describes the detailed procedures used for field data acquisition on post-tensioned concrete bridges, with the goal of providing a transparent and replicable experimental framework.

\subsection{Acquisition Setup}
\label{section 4.1}
\begin{figure}[!ht]
    \centering
    \begin{subfigure}[b]{0.9\textwidth}
        \centering
        \includegraphics[width=0.95\linewidth]{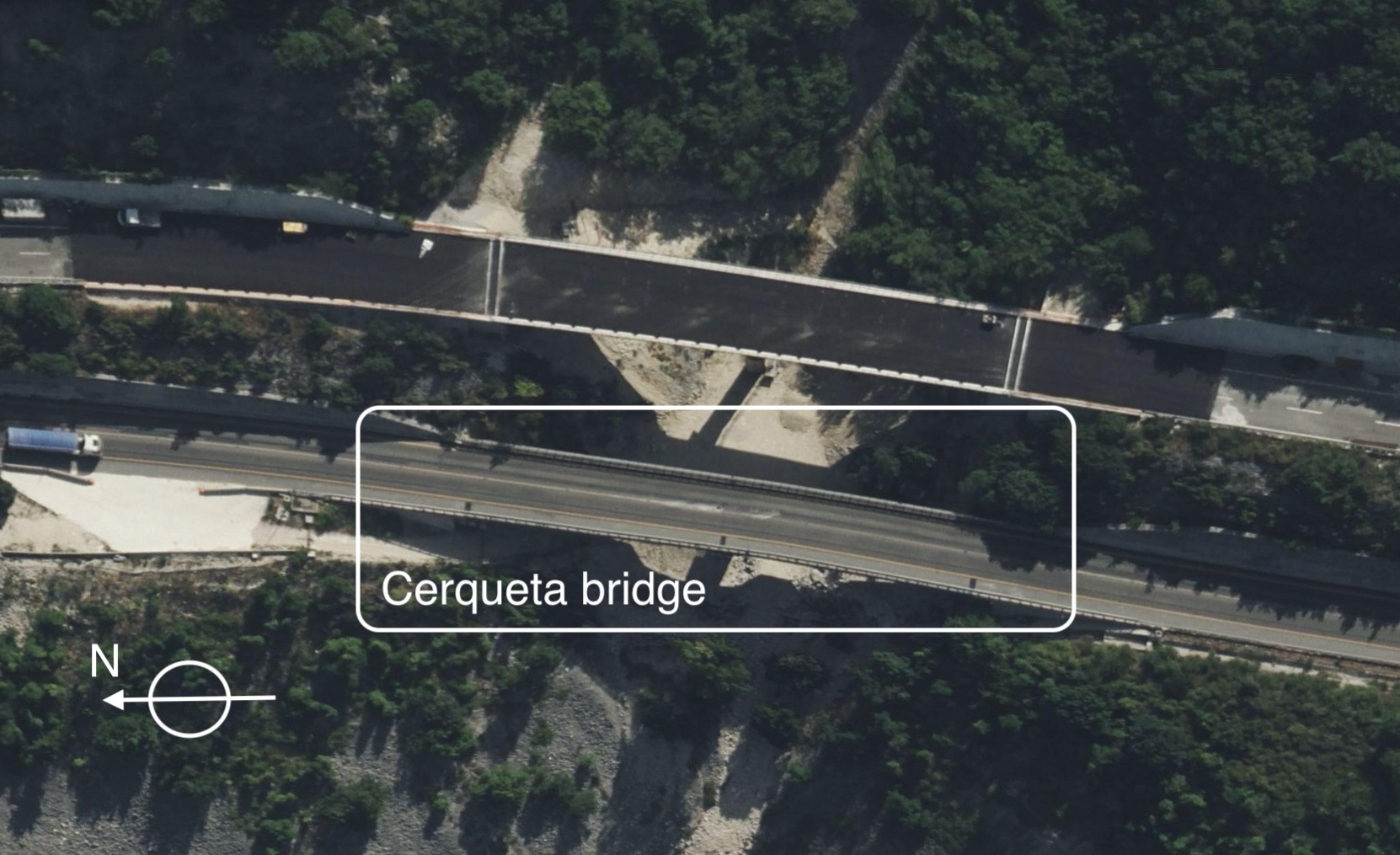}
        \caption*{(a)}
    \end{subfigure}
    \hfill
    
    \begin{subfigure}[b]{0.9\textwidth}
        \centering
        \includegraphics[width=0.95\linewidth]{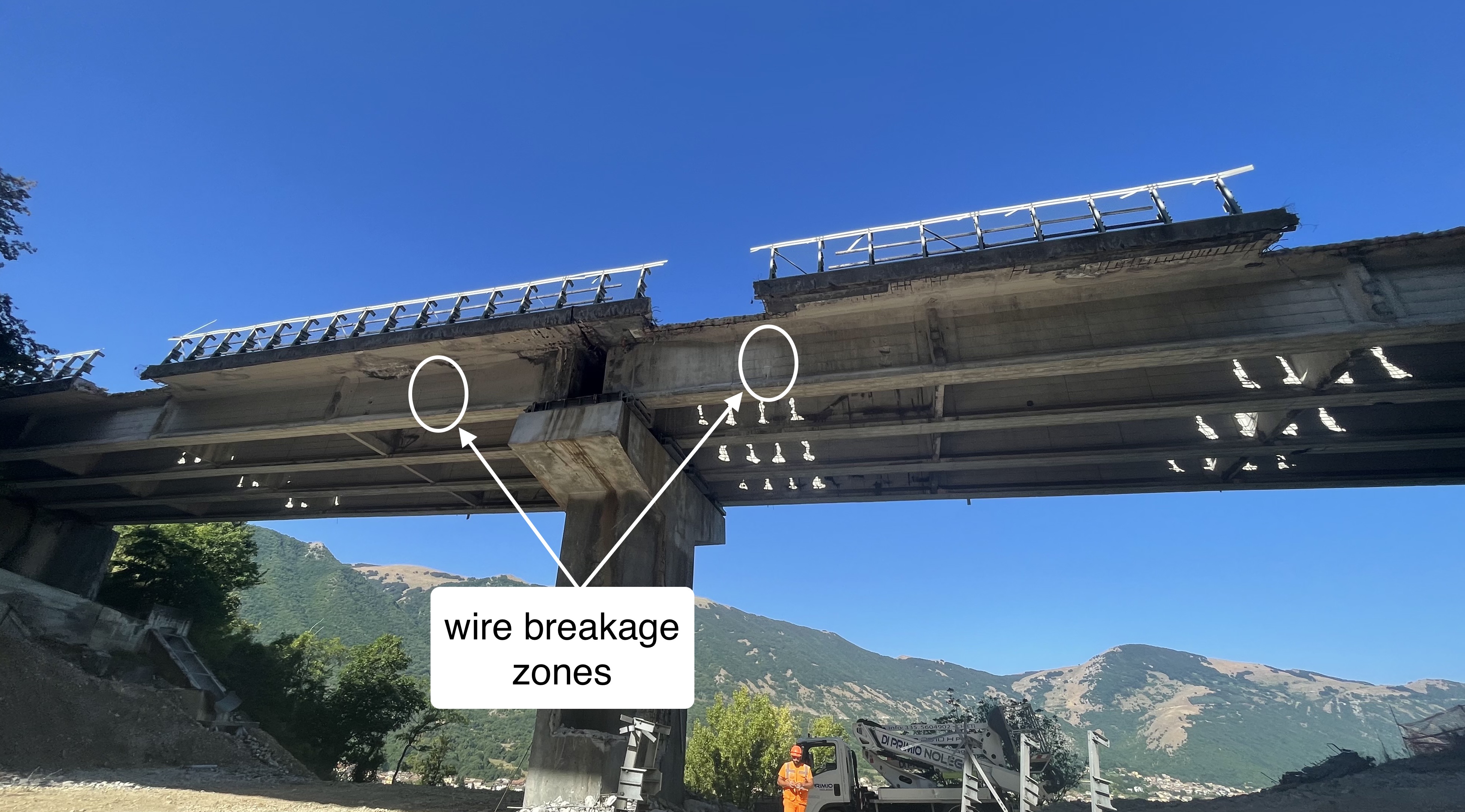}
        \caption*{(b)}
    \end{subfigure}
    \caption{Cerqueta bridge along the A24 highway selected for controlled prestressing tendon breakage tests: (a) plan view; (b) lateral view from East.}

    \label{fig:acc_bridges}
\end{figure}

The experimental framework was conducted on the post-tensioned concrete bridge Cerqueta, located along the A24 highway in L'Aquila, Italy. As the structure was scheduled for demolition, it offered a rare opportunity to perform controlled breakage tests on its internal prestressing tendons (Fig. \ref{fig:acc_bridges}).
Cerqueta is a two-span bridge, each span consisting of four longitudinal prestressed reinforced
concrete I-girders with a cast-inplace deck slab; the longitudinal girders are connected by cross beams. I-girders are prestressed by post-tensioned cables, placed in grouted corrugated metallic ducts. The prestressing system of the I-beams consisted of seven tendons with parabolic path having 12 strands with seven twisted wires each.
A detailed summary of the structural and experimental parameters for the bridge is provided in Table~\ref{tab:bridge_params}.

%

\begin{table}[!ht]
    \centering
    \caption{Summary of structural and experimental parameters for Cerqueta bridge.}
    \label{tab:bridge_params}
    \scriptsize
    \renewcommand{\arraystretch}{1.5}
    \begin{tabularx}{0.62\textwidth}{@{}lX@{}}
        \toprule
        \textbf{Parameter} & \textbf{Cerqueta Bridge} \\
        \midrule
        Location & A24 Highway, L'Aquila \\
        Span Length & 38 m \\
        Beam height & 2.12 m \\
        Orientation & North–South longitudinal axis \\
        Test Side & East-facing side beam \\
        Height above ground & ~15 m \\
        Number of Tendons Monitored & 4 (12 strands each, with twisted wires) \\
        Accelerometers & 2 (4.5 m from cut area) \\
        Sampling Rate & 96 kHz \\
        Data Acquisition Mode & Dual-channel, synchronized \\
        \bottomrule
    \end{tabularx}
\end{table}

\begin{figure}[!ht]
    \centering
    \begin{subfigure}[b]{0.48\textwidth}
        \centering
        \includegraphics[width=0.95\textwidth]{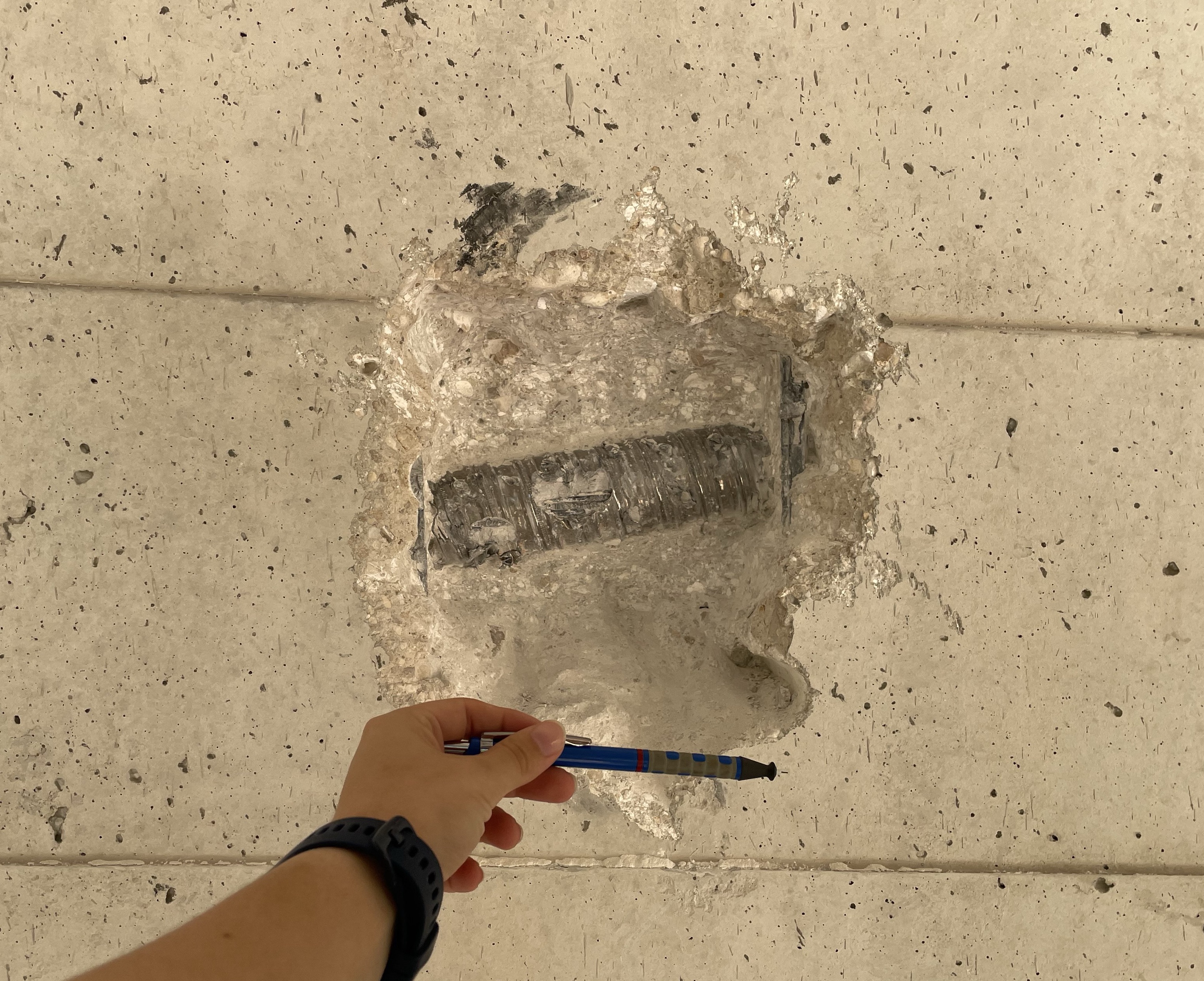}
        \caption*{(a)}
    \end{subfigure}
    \hfill
    \begin{subfigure}[b]{0.48\textwidth}
        \centering
        \includegraphics[width=0.95\textwidth]{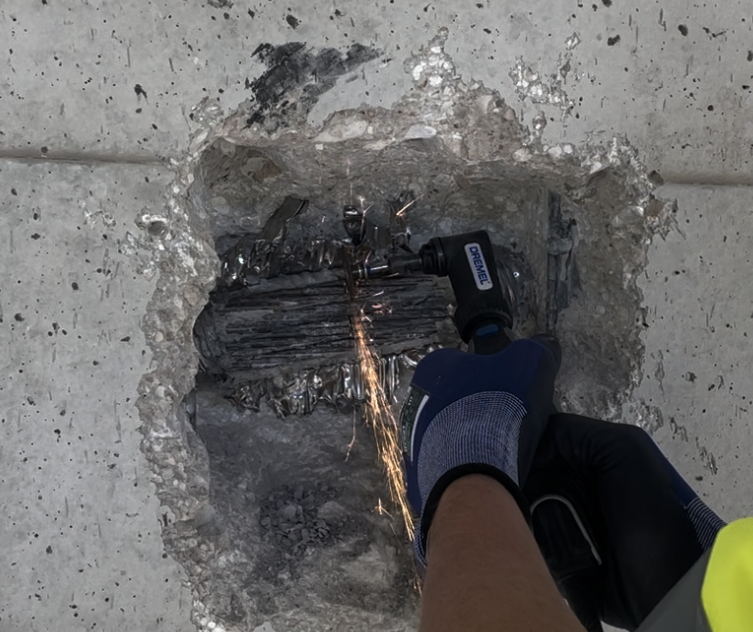}
        \caption*{(b)}
    \end{subfigure}
    \begin{subfigure}[b]{0.48\textwidth}
        \centering
        \includegraphics[width=0.85\textwidth, angle=-90]{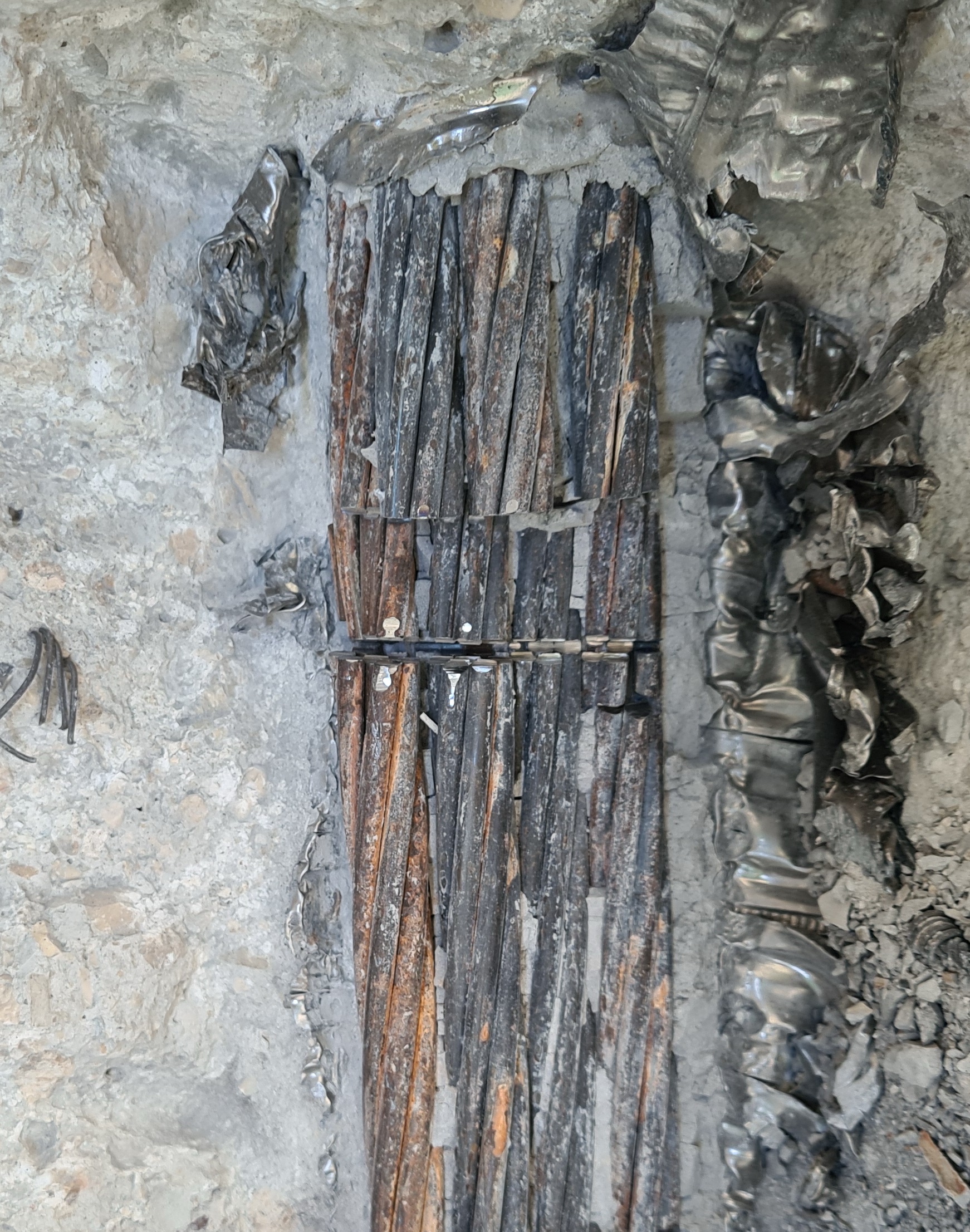}
        \caption*{(c)}
    \end{subfigure}
    
    \caption{Controlled wire breakage test on the Cerqueta bridge: (a) operational window opened in the beam's web; (b) wire cutting using an electric trimmer; (c) strands with twisted wires after breakage}

    \label{fig:twisted_tendon}
\end{figure}

These tests were made possible through a collaboration between Politecnico di Torino and Strada dei Parchi S.p.A., which allowed data collection during the partial deconstruction phase. Due to the considerable height and limited accessibility of the bridge, a Mobile Elevated Work Platform (MEWP) was utilized to access the side beams where instrumentation and testing were performed. The experimental procedure began by carefully removing sections of concrete in the web of the I-beams to open an operational window to access the internal steel ducts housing the prestressing tendons. In total, four windows were opened, two for each of east-facing side beams of the two spans (see Fig. \ref{fig:twisted_tendon}). Controlled wire breakage was performed by cutting the cross section of wires with an electric trimmer tool until generating a spontaneous tensile breaking. Considering the actual stress in the wires, they break when about half the cross section is cut.

%

\subsection{Data Acquisition}
\label{section 4.2}
\begin{figure}[!ht]
    \centering
    \begin{subfigure}[b]{0.9\textwidth}
        \centering
        \includegraphics[width=0.7\linewidth]{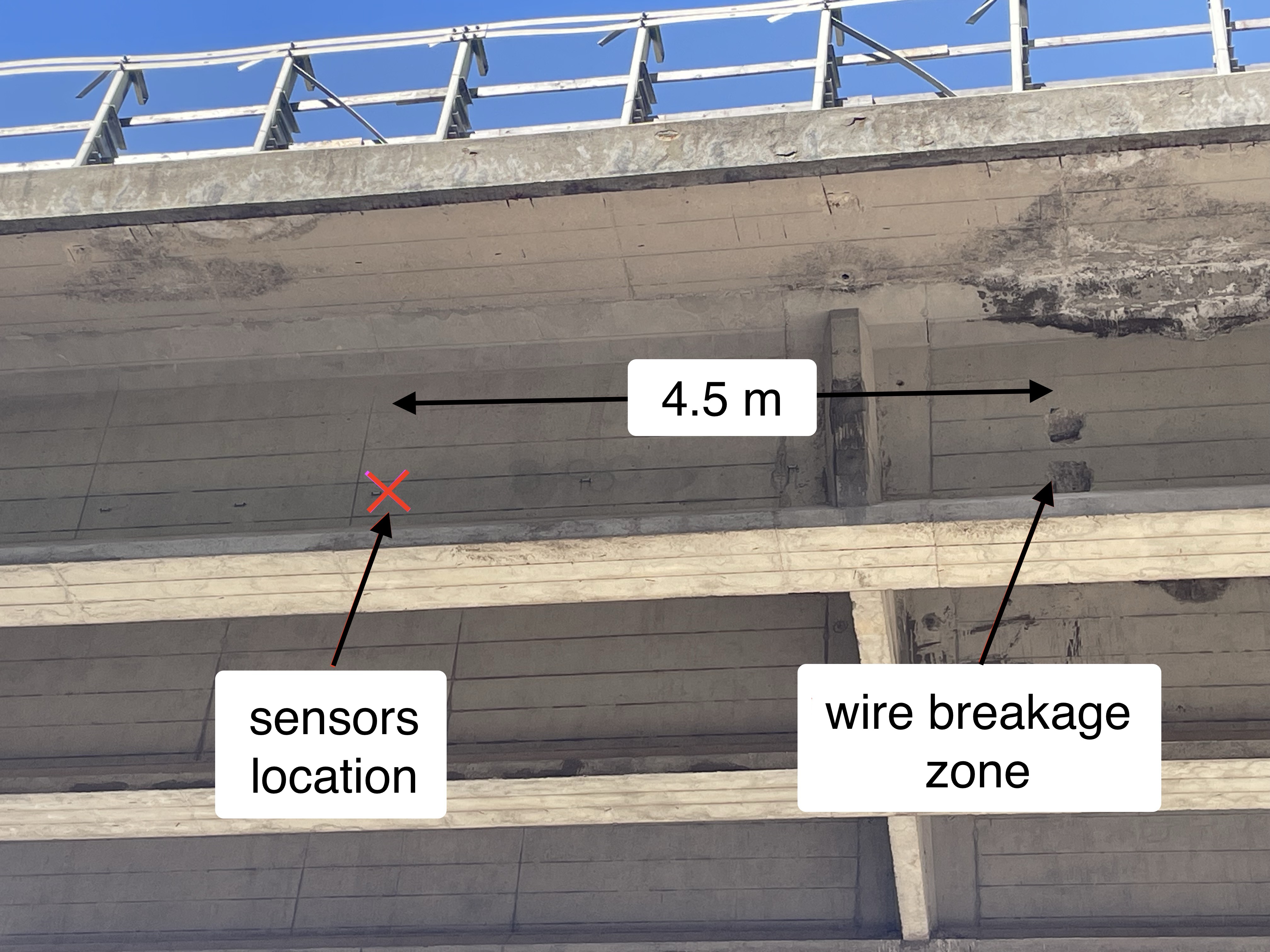}
        \caption*{(a)}
    \end{subfigure}\\
    \begin{subfigure}[b]{0.9\textwidth}
        \centering
        \includegraphics[width=0.4\linewidth, angle=-90]{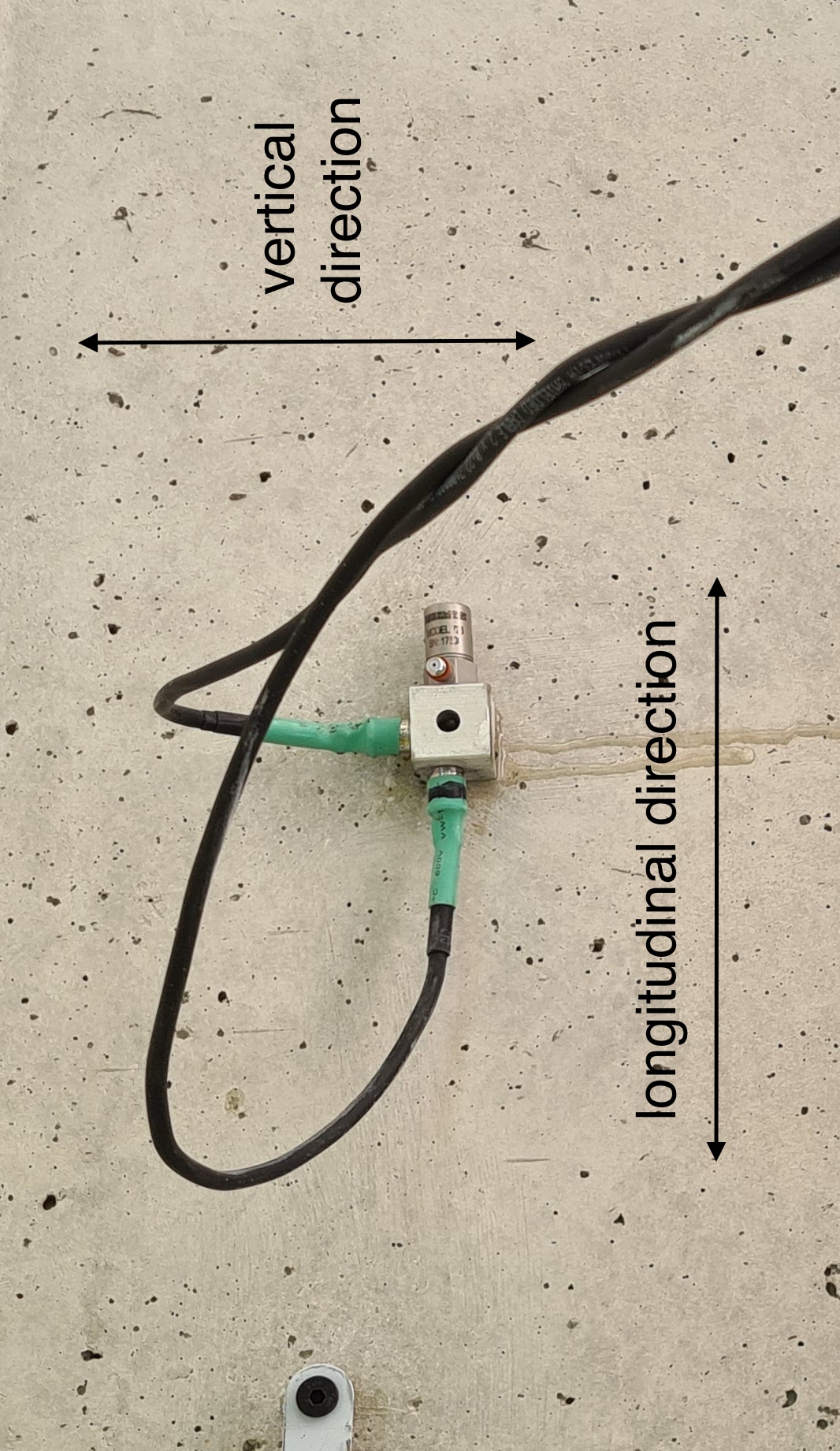}
        \caption*{(b)}
    \end{subfigure}
    \caption{Sensor positioning: (a) distance from the cutting areas; (b) orientation of the two accelerometers (green devices) with respect to the bridge alignment}
    \label{fig:acc_sensors}
\end{figure}

The acquisition system included two piezo-ceramic mono-axial accelerometers (Model 805M1, DSPM Industria srl) having frequency response from 0.4 Hz to 12 kHz, and natural frequency of 34 kHz. Both accelerometers were mounted on the web of the side beam where the wires were cut, 4.5 m far from the cutting area, towards the midspan. The two accelerometers were installed to catch vibrations in the longitudinal direction and in the vertical direction, with respect to the bridge alignment (Fig. \ref{fig:acc_sensors}).
Although conventional structural monitoring typically targets low frequencies, up to about 50 Hz, to study the natural frequencies of the bridge, in our sound event detection based approach, the accelerometers are used as microphones to capture the acoustic events propagating through the structure. That is the reason for the selected frequency range of the accelerometers. Consistently, an acquisition rate typical of sound recorders has been set, equal to 96 kHz. Besides a technical consistency of the devices with the developed approach, a key feature of the acquisition system is its simplicity and low cost.
Data were recorded in continuous via a dual-channel acquisition system labeled as left and right channels, each capturing different orientations of acoustic signals. This setup provided a more detailed view of the events, enabling a richer analysis of time-frequency characteristics and signal propagation.

%

%

In addition to wire breakage acoustic signals, two other classes of acoustic events were recorded on the Cerqueta bridge, namely electric trimmer operation and light hammering. Furthermore, to further enrich the dataset, traffic noise was recorded, using the same acquisition system, on a bridge of the same type as Cerqueta. All in all, this experimental activity ensures that the dataset reflects a realistic scenario during structural monitoring of post-tensioned systems, enabling the development of more generalized and high-performing SHM systems. It is also important to highlight that acoustic data from each sensor were modeled independently, with each uniaxial signal treated as a separate input for the generative model. This decision was made to increase the number of input signals and its diversity for the models to avoid bias.

\section{Results and Discussion}
\label{sec5}

\subsection{Quantitative Results}
This section presents a quantitative evaluation of the proposed GAN models for generating STFT spectrograms. The objective is to assess how well the generated samples preserve the structural and statistical characteristics of real spectrograms. Three widely used metrics were adopted: Structural Similarity Index (SSIM), Peak Signal-to-Noise Ratio (PSNR), and Fréchet Inception Distance (FID). Table \ref{tab:gan_results} summarizes the performance of all models across different event classes and STFT window sizes (128 and 256).

\begin{table*}[htb]
    \centering
    \caption{Performance metrics of GAN variants for generating STFT spectrograms.}
    \label{tab:gan_results}
    \scriptsize
    \renewcommand{\arraystretch}{1.5} 
    \begin{tabularx}{0.87\textwidth}{@{}llllllllll@{}}
    \toprule
        \multirow{2}{*}{\textbf{Metrics}} 
        & \multirow{2}{*}{\textbf{Class}} 
        & \multicolumn{2}{c}{\textbf{DCGAN}} 
        & \multicolumn{2}{c}{\textbf{WGAN-GP}} 
        & \multicolumn{2}{c}{\textbf{LSGAN}} 
        & \multicolumn{2}{c}{\textbf{STFTSynth}} \\ 
        \cmidrule(lr){3-4} \cmidrule(lr){5-6} \cmidrule(lr){7-8} \cmidrule(lr){9-10}
        & & \textbf{128} & \textbf{256} & \textbf{128} & \textbf{256} & \textbf{128} & \textbf{256} & \textbf{128} & \textbf{256} \\
        \midrule

        
        \multirow{4}{*}{\textbf{SSIM}} 
        & \textbf{Breakage} & 0.358 & 0.325 & 0.367 & 0.180 & 0.360 & 0.185 & \textbf{0.488} & \textbf{0.459} \\
        & \textbf{Trimmer} & 0.336 & 0.393 & 0.342 & 0.403 & 0.347 & 0.404 & \textbf{0.456} & \textbf{0.532} \\
        & \textbf{Hammer} & 0.222 & 0.225 & 0.208 & 0.232 & 0.212 & 0.224 & \textbf{0.326} & \textbf{0.287} \\
        & \textbf{Traffic} & 0.189 & 0.264 & 0.170 & 0.186 & 0.176 & 0.184 & \textbf{0.265} & \textbf{0.295} \\
        \midrule

        \multirow{4}{*}{\textbf{PSNR}} 
        & \textbf{Breakage} & 13.348 & 13.451 & 13.409 & 11.972 & 13.523 & 12.249 & \textbf{13.731} & \textbf{14.009} \\
        & \textbf{Trimmer} & 12.772 & 13.258 & \textbf{13.424} & 13.621 & 13.284 & 13.719 & \textbf{13.422} & \textbf{14.302} \\
        & \textbf{Hammer} & 10.956 & 11.427 & 10.520 & 11.289 & 10.689 & 11.393 & \textbf{11.338} & \textbf{11.934} \\
        & \textbf{Traffic} & 11.521 & 11.530 & 11.374 & 12.020 & 11.928 & 11.957 & \textbf{12.422} & \textbf{12.232} \\
        \midrule

        \multirow{4}{*}{\textbf{FID}} 
        & \textbf{Breakage} & 0.253 & \textbf{0.229} & 0.219 & 0.331 & 0.274 & 0.231 & \textbf{0.210} & 0.249 \\
        & \textbf{Trimmer} & 0.251 & 0.242 & \textbf{0.173} & 0.270 & 0.186 & \textbf{0.190} & 0.219 & 0.235 \\
        & \textbf{Hammer} & 0.209 & 0.177 & 0.221 & 0.194 & 0.237 & 0.241 & \textbf{0.102} & \textbf{0.117} \\
        & \textbf{Traffic} & 0.209 & 0.279 & \textbf{0.179} & 0.237 & 0.198 & \textbf{0.167} & 0.249 & 0.246 \\
    \bottomrule
    \end{tabularx}
\end{table*}

SSIM evaluates the perceptual similarity between generated and real STFT spectrograms, where higher values indicate better performance. The results in Table \ref{tab:gan_results} show that STFTSynth outperforms all other models across all events and window sizes, achieving the highest SSIM scores. For instance, STFTSynth achieves an SSIM of 0.488 for the Breakage event (window size 128) and 0.532 for the Trimmer event (window size 256), both of which are the highest recorded values. This indicates that the proposed STFTSynth model generates STFT spectrograms with better similarity to real spectrograms, effectively preserving key spectral features. Contrarily, WGAN-GP exhibits the lowest SSIM scores across almost all event classes, particularly for the Breakage event (window size 256), with a score of 0.180, and the Traffic event (window size 128), with a score of 0.170. This highlights that WGAN-GP struggles to preserve the fine-grained structural details present in real STFT spectrograms.

PSNR measures reconstruction quality by quantifying the distortion between generated and real spectrograms, where higher values indicate lower distortion and improved clarity. Similar to SSIM, STFTSynth demonstrates the best overall performance, achieving the highest PSNR scores across all event classes and window sizes. The Trimmer event obtained a score of 14.302, and the Breakage event achieved a score of 14.009, both using a window size of 256. These results show that STFTSynth generates high-quality spectrograms with minimal distortion. In contrast, WGAN-GP consistently produces lower PSNR values, particularly for the Hammer event using a window size of 128, obtaining a score of 10.520, which indicates higher noise levels and lower clarity in the generated STFT spectrograms. However, this model achieves comparable results to STFTSynth for the Trimmer event using a window size of 128. The other two models, DCGAN and LSGAN, exhibit moderate performance compared to STFTSynth.

FID measures the similarity between real and generated data distributions by comparing their feature representations. Lower FID values indicate that generated spectrograms are more realistic and closely resemble real STFT data. As shown in Table \ref{tab:gan_results}, STFTSynth generally achieves the lowest FID scores, notably for the Breakage event (0.210, window size 128) and Hammer events (0.102 and 0.117, window sizes 128 and 256, respectively). These results confirm STFTSynth’s capability to generate spectrograms closely aligned with real data distributions. However, STFTSynth’s performance is not uniformly superior in all cases. For the Breakage event at window size 256, DCGAN achieves a slightly lower FID (0.229) compared to STFTSynth’s 0.249, demonstrating DCGAN’s competitiveness in specific instances. Additionally, for the Trimmer event at window size 256, LSGAN (0.190) outperforms STFTSynth (0.235), suggesting LSGAN’s ability to better capture realistic STFT spectrogram features under certain conditions.

Moreover, WGAN-GP also shows competitive performance in some cases. Specifically, for the Trimmer (0.173) and Traffic (0.179) events at a window size of 128, WGAN-GP achieves the lowest FID, surpassing both LSGAN and STFTSynth. Despite this strength, WGAN-GP demonstrates inconsistent performance, with significantly higher FID scores for other events such as Breakage and Hammer. These variations suggest that WGAN-GP struggles to consistently capture high-quality spectrogram features across diverse event types. Lastly, DCGAN typically produces higher FID scores compared to the other evaluated models, indicating limitations in accurately modeling complex structural details and capturing realistic data distributions.

\subsection{Qualitative Analysis}
\begin{figure}[htb]
    \centering
    \begin{tikzpicture}
        \node[anchor=south west, inner sep=0] (image) at (0,0) {
            \begin{minipage}{\textwidth}
                \centering
                \begin{subfigure}[b]{0.19\textwidth}
                    \centering
                    \includegraphics[width=\textwidth]{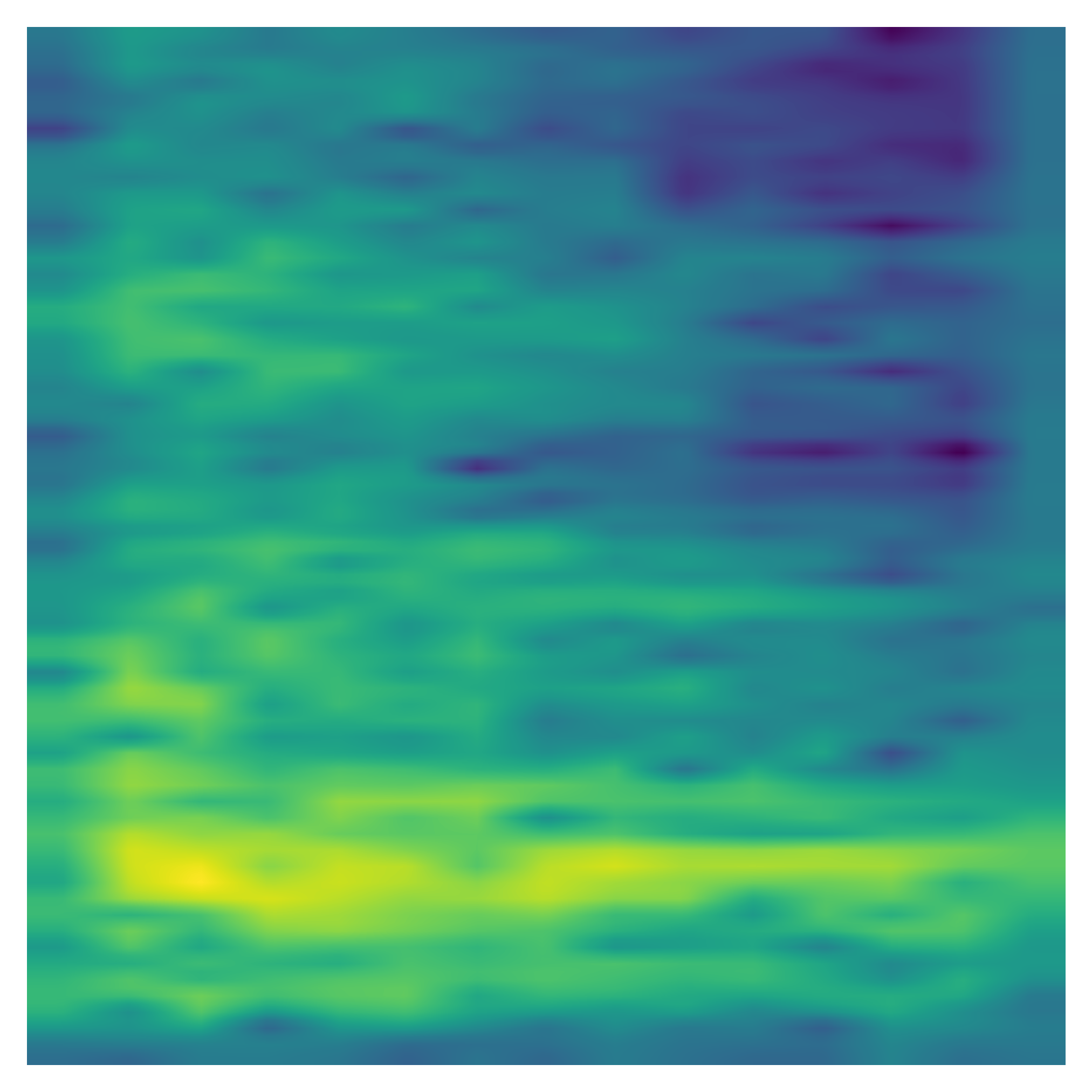}
                    \caption*{(a)}
                \end{subfigure}
                \hfill
                \begin{subfigure}[b]{0.19\textwidth}
                    \centering
                    \includegraphics[width=\textwidth]{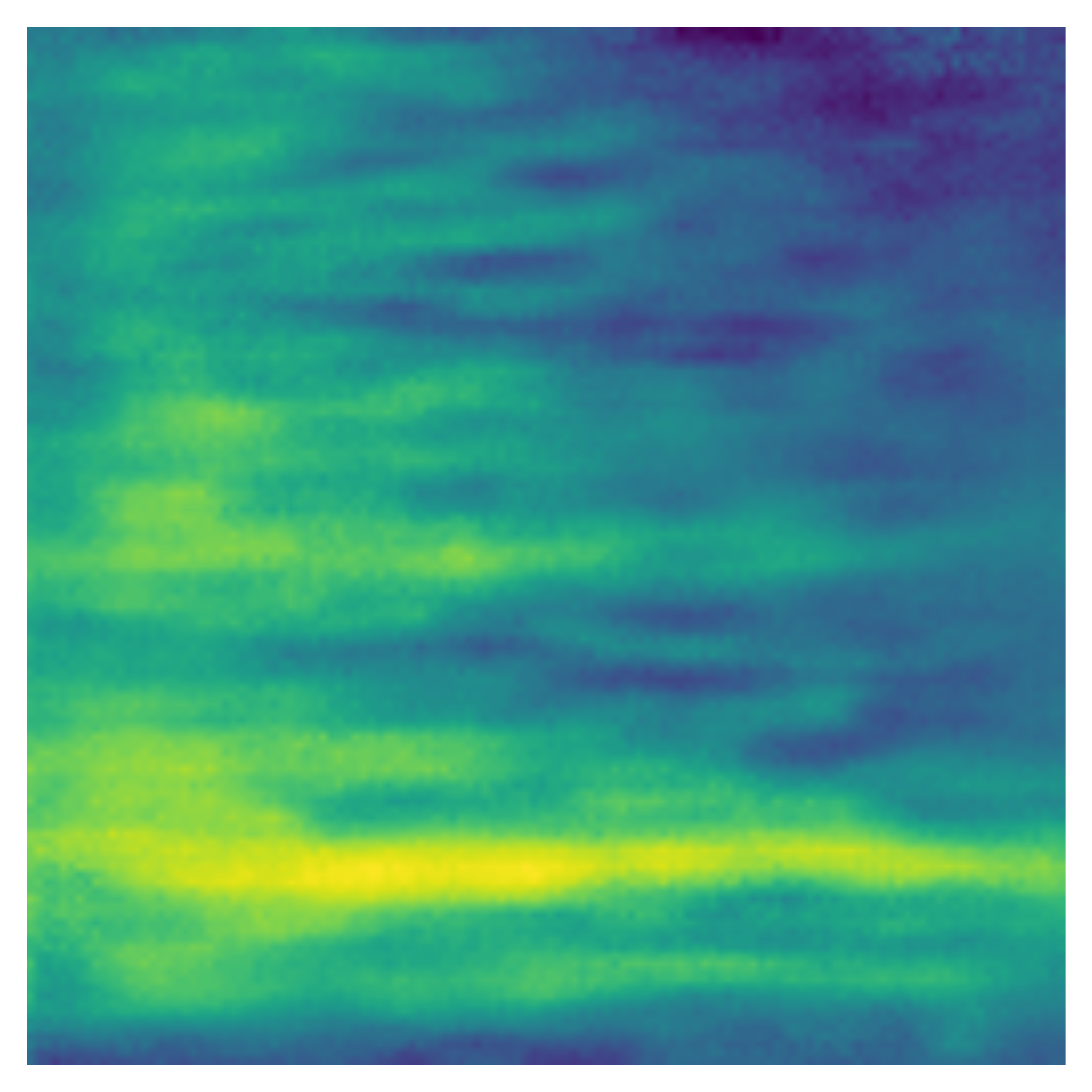}
                    \caption*{(b)}
                \end{subfigure}
                \hfill
                \begin{subfigure}[b]{0.19\textwidth}
                    \centering
                    \includegraphics[width=\textwidth]{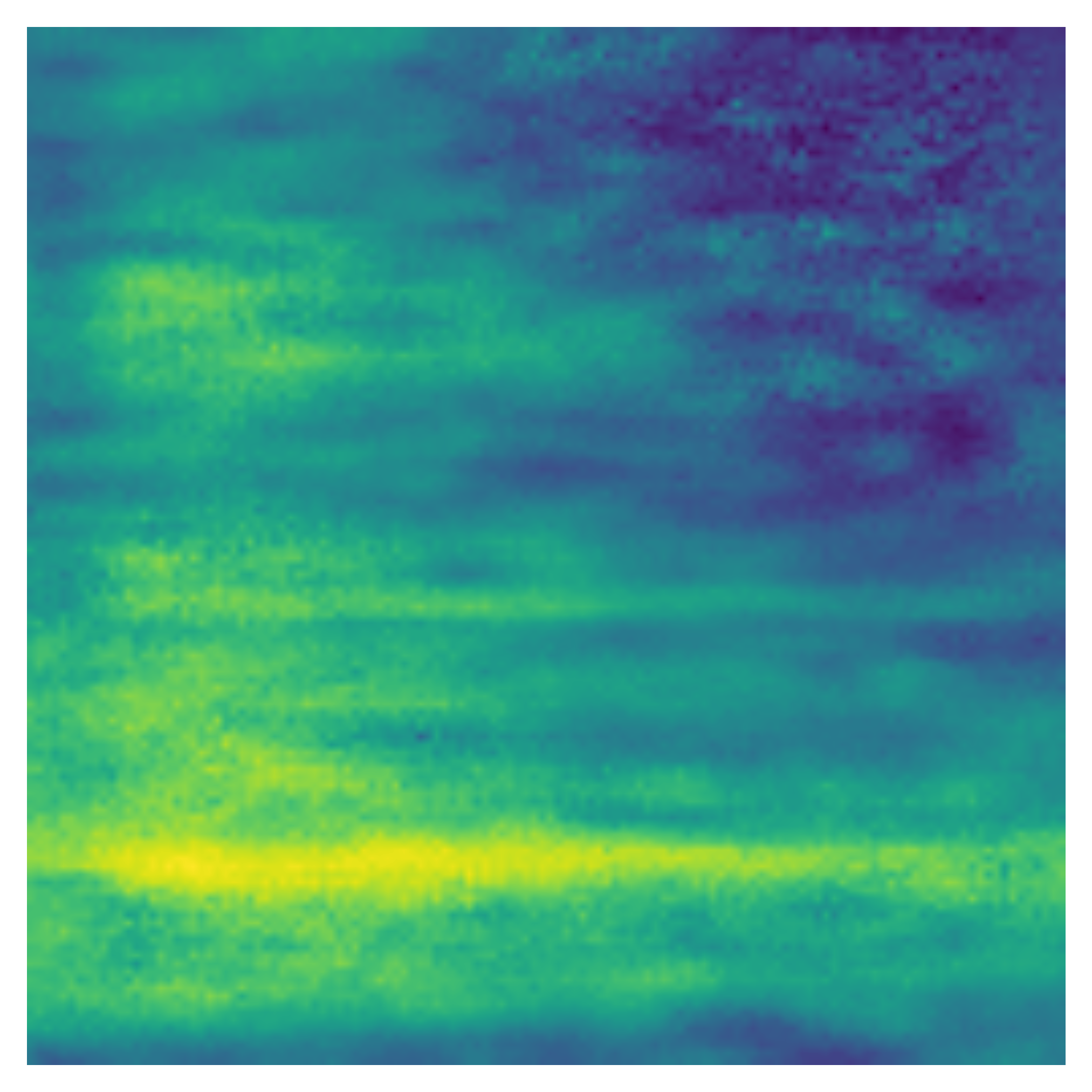}
                    \caption*{(c)}
                \end{subfigure}
                \begin{subfigure}[b]{0.19\textwidth}
                    \centering
                    \includegraphics[width=\textwidth]{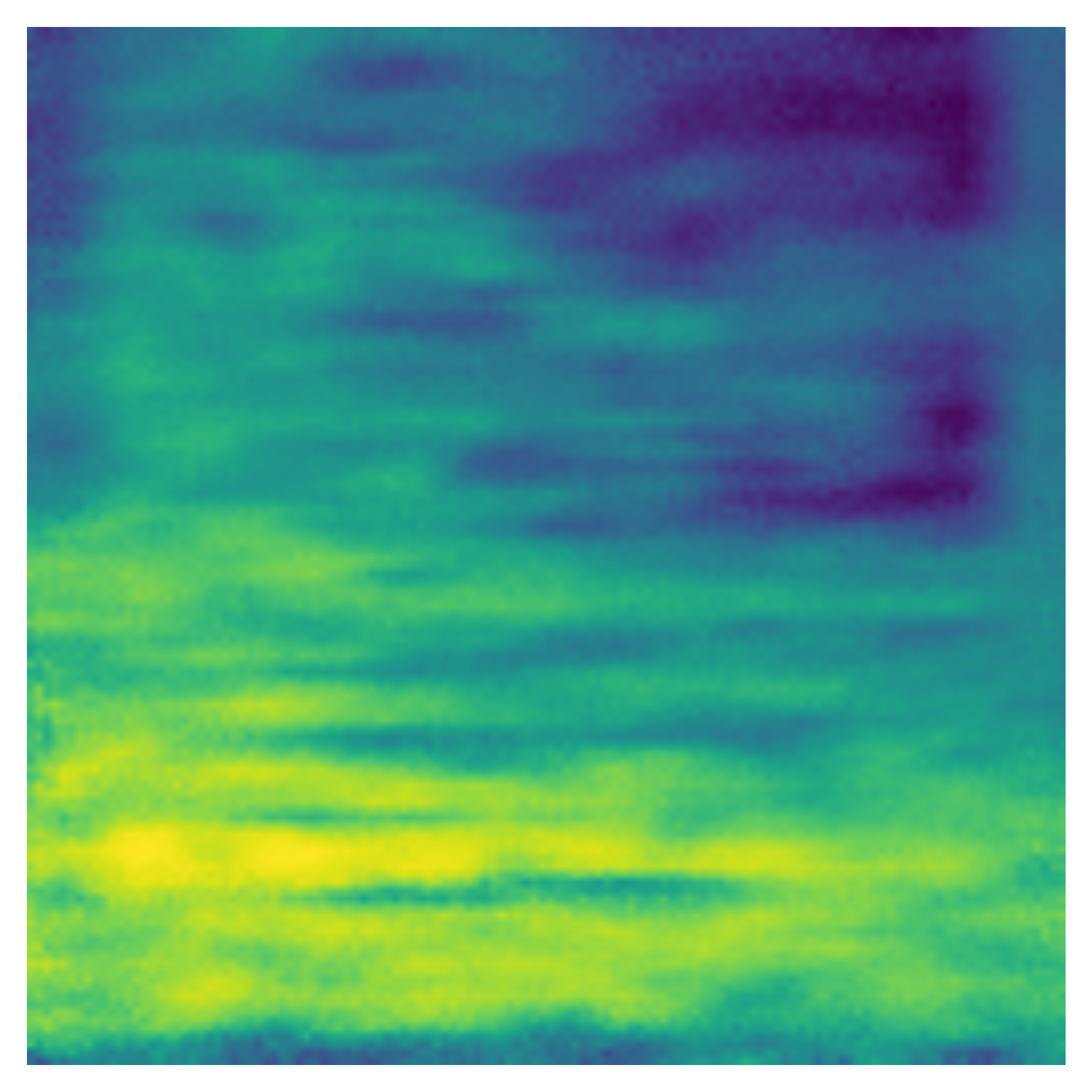}
                    \caption*{(d)}
                \end{subfigure}
                \hfill
                \begin{subfigure}[b]{0.19\textwidth}
                    \centering
                    \includegraphics[width=\textwidth]{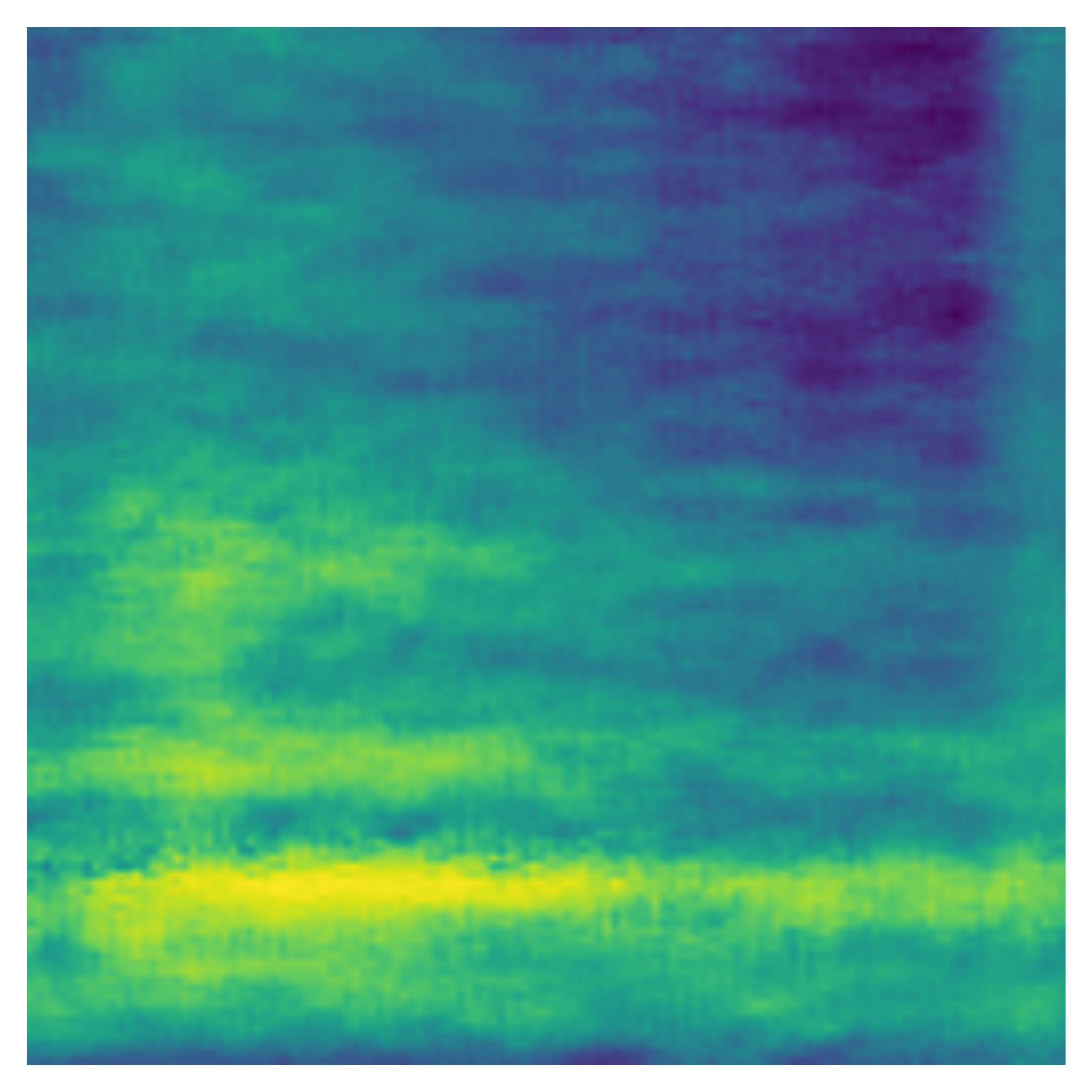}
                    \caption*{(e)}
                \end{subfigure}
            \end{minipage}
        };

        \begin{scope}[x={(image.south east)}, y={(image.north west)}]
            \draw[->, thick] (0.005, 0.15) -- (0.06, 0.15); 
            \node at (0.015, 0) {\footnotesize \textbf{Time (s)}};

            \draw[->, thick] (-0.005, 0.15) -- (-0.005, 0.35); 
            \node[rotate=90] at (-0.025, 0.6) {\footnotesize \textbf{Frequency (Hz)}};
        \end{scope}
    \end{tikzpicture}

    \caption{STFT representations of real and generated wire breakage spectrograms (window size = 128): (a) Real STFT, (b) DCGAN, (c) WGAN, (d) LSGAN, and (e) STFTSynth [Full Model]. The horizontal and vertical arrows indicate the time and frequency axes, respectively}
    \label{fig:stft_gan_128}
\end{figure}

\begin{figure}[htb]
    \centering
    \begin{tikzpicture}
        \node[anchor=south west, inner sep=0] (image) at (0,0) {
            \begin{minipage}{\textwidth}
                \centering
                \begin{subfigure}[b]{0.19\textwidth}
                    \centering
                    \includegraphics[width=\textwidth]{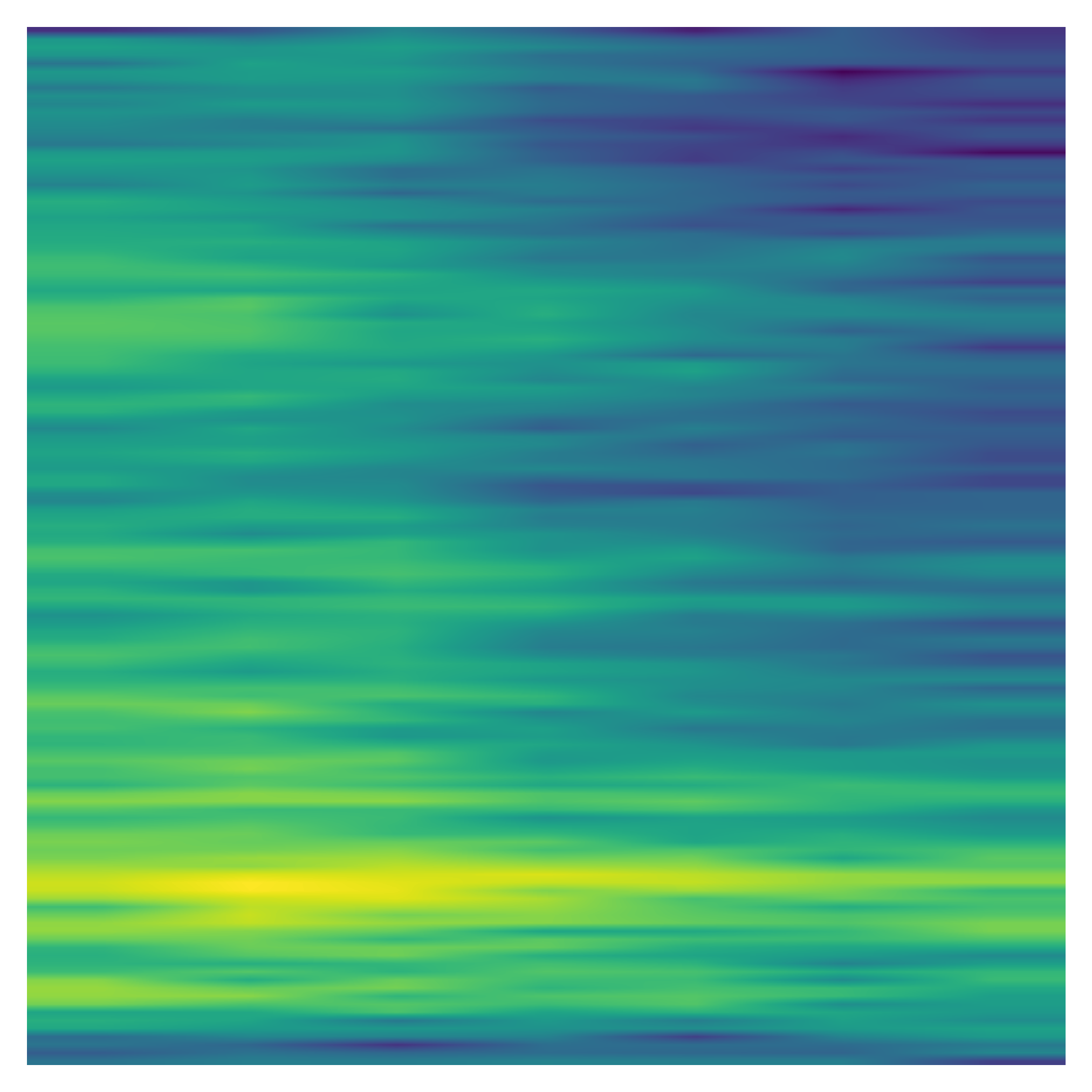}
                    \caption*{(a)}
                \end{subfigure}
                \hfill
                \begin{subfigure}[b]{0.19\textwidth}
                    \centering
                    \includegraphics[width=\textwidth]{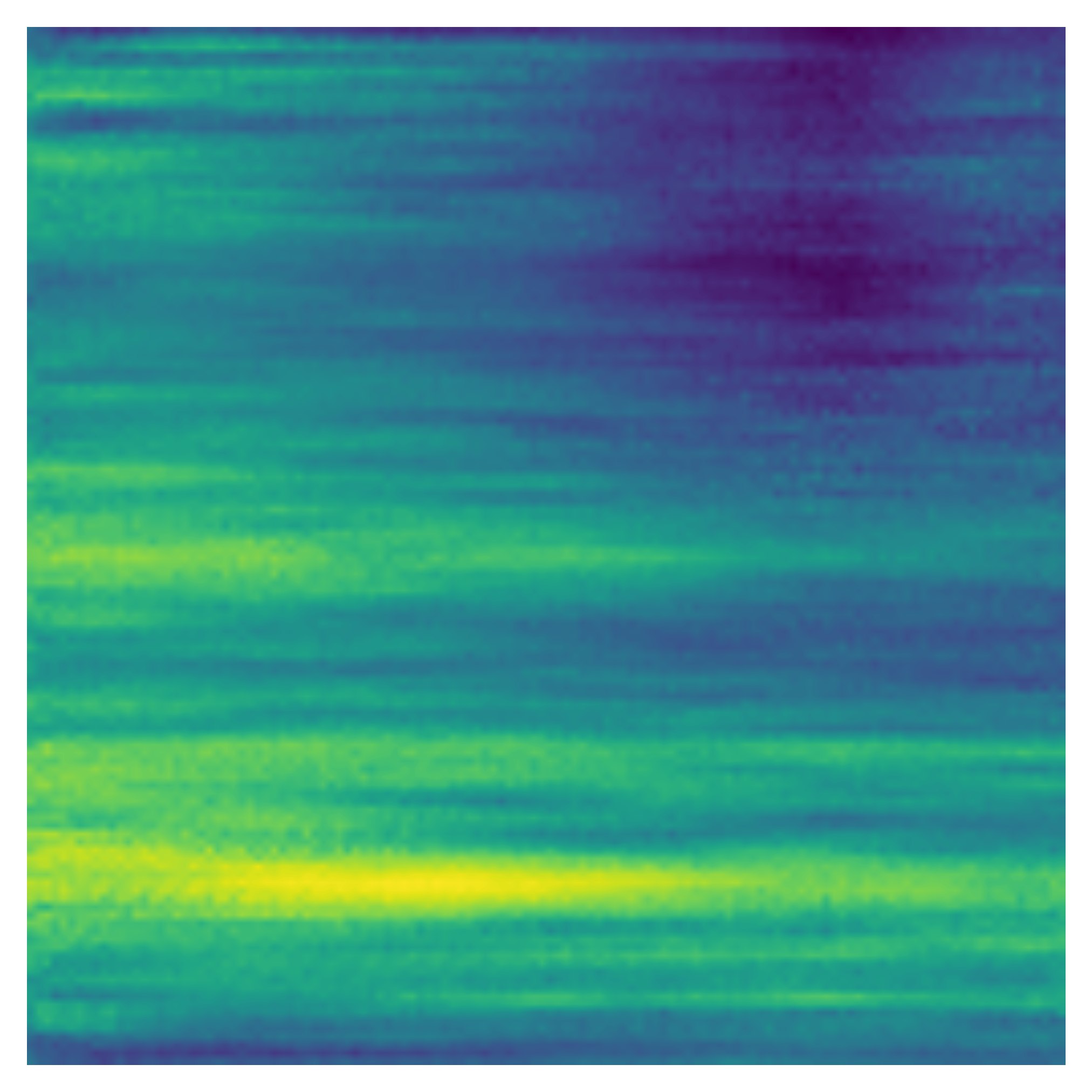}
                    \caption*{(b)}
                \end{subfigure}
                \hfill
                \begin{subfigure}[b]{0.19\textwidth}
                    \centering
                    \includegraphics[width=\textwidth]{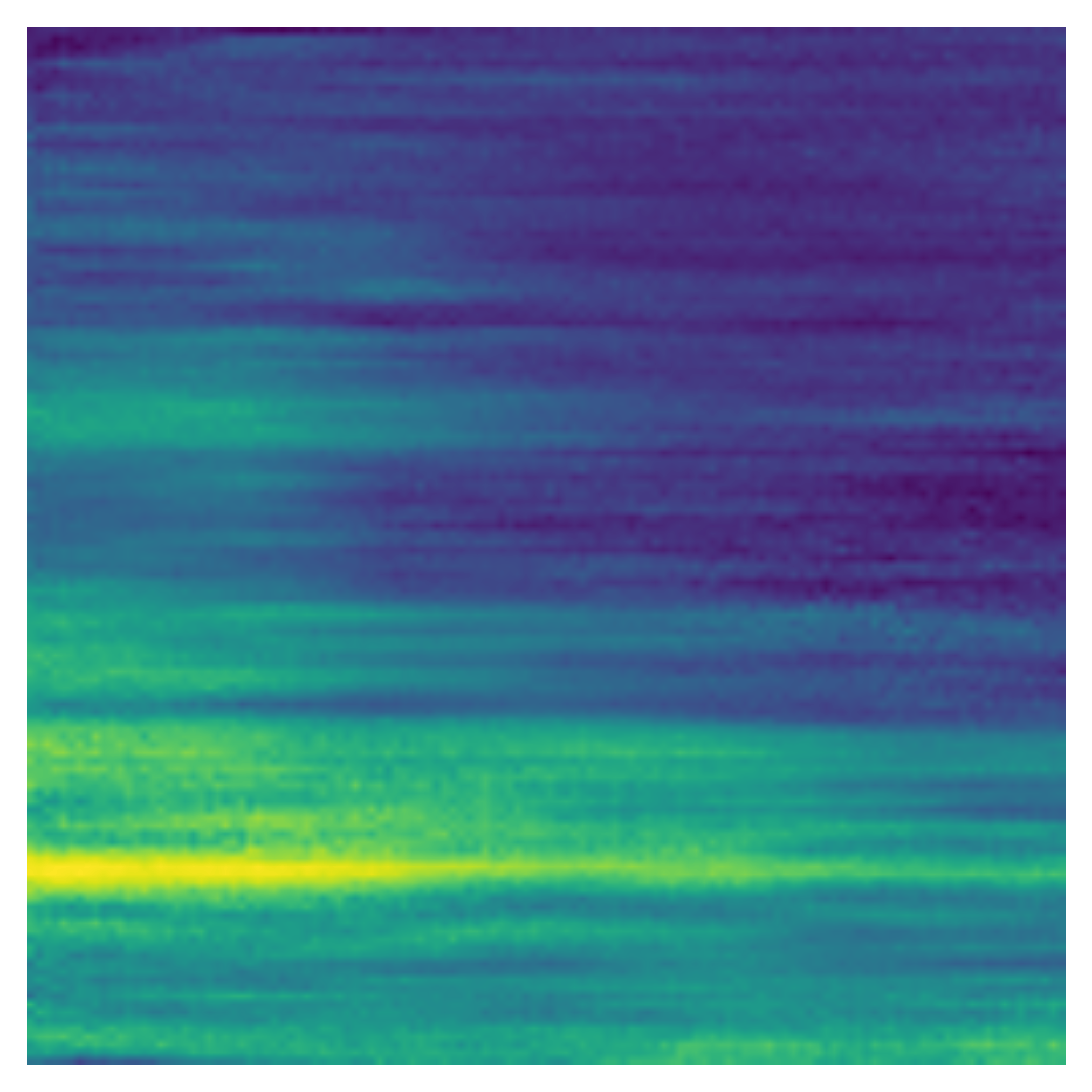}
                    \caption*{(c)}
                \end{subfigure}
                \begin{subfigure}[b]{0.19\textwidth}
                    \centering
                    \includegraphics[width=\textwidth]{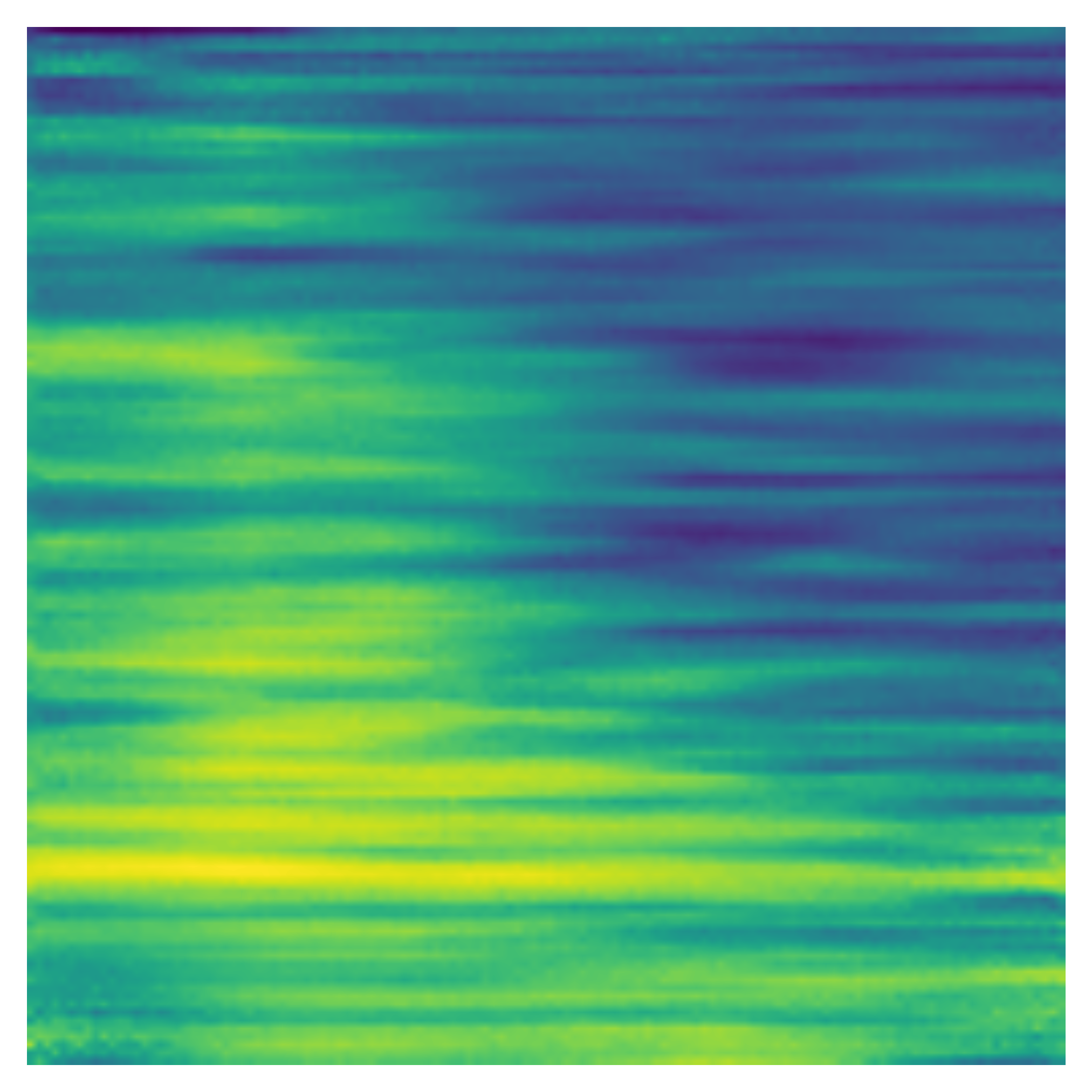}
                    \caption*{(d)}
                \end{subfigure}
                \hfill
                \begin{subfigure}[b]{0.19\textwidth}
                    \centering
                    \includegraphics[width=\textwidth]{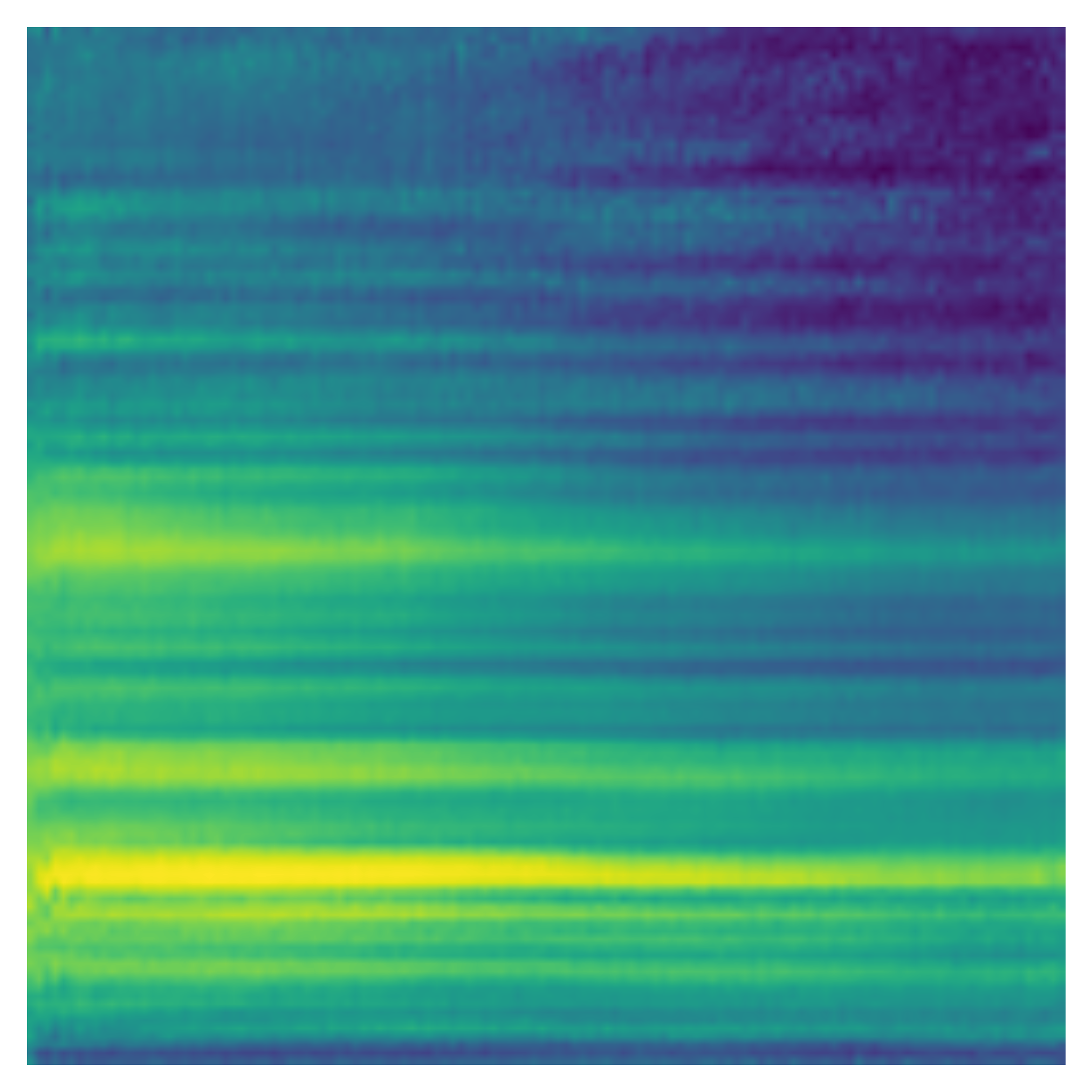}
                    \caption*{(e)}
                \end{subfigure}
            \end{minipage}
        };

        \begin{scope}[x={(image.south east)}, y={(image.north west)}]
            \draw[->, thick] (0.005, 0.15) -- (0.06, 0.15); 
            \node at (0.015, 0) {\footnotesize \textbf{Time (s)}};

            \draw[->, thick] (-0.005, 0.15) -- (-0.005, 0.35); 
            \node[rotate=90] at (-0.025, 0.6) {\footnotesize \textbf{Frequency (Hz)}};
        \end{scope}
    \end{tikzpicture}

    \caption{STFT representations of real and generated wire breakage spectrograms (window size = 256): (a) Real STFT, (b) DCGAN, (c) WGAN, (d) LSGAN, and (e) STFTSynth [Full Model]. The horizontal and vertical arrows indicate the time and frequency axes, respectively}
    \label{fig:stft_gan_256}
\end{figure}

In addition to the quantitative evaluation, a qualitative visual assessment of the generated STFT spectrograms was performed. The generated STFT spectrograms were visually compared against real ones to examine how accurately each model captures the key time-frequency characteristics of real events. Figures \ref{fig:stft_gan_128} and \ref{fig:stft_gan_256} illustrate the real and generated STFT spectrograms for wire breakage across the different GAN models, using STFT window sizes of 128 and 256, respectively. A visual inspection confirms that STFTSynth consistently produces the sharpest and most structurally detailed spectrograms, closely matching the real-world patterns. Among the baseline models, LSGAN generates spectrograms closest in quality to STFTSynth. In contrast, DCGAN and WGAN-GP show notable limitations, particularly in silent or less-structured regions, resulting in visible distortions or missing spectral details.

It is important to highlight that the evaluation metrics used in this study assess the quality of the generated STFT spectrograms at the representation level. Improvements in SSIM and PSNR indicate better preservation of spectral patterns, while FID quantifies the similarity between the distributions of real and generated samples. These metrics provide evidence that the synthetic spectrograms capture the key statistical and structural characteristics of the original data. However, they do not directly measure performance in downstream SHM tasks such as event classification or detection. Therefore, they should be interpreted as necessary but not sufficient indicators of the potential usefulness of synthetic data in subsequent monitoring applications.

\subsection{Ablation Study}
\label{sec:ablation}
An ablation study was conducted to investigate the individual contributions of the DRBs and BiGRU layers within the STFTSynth architecture. Different variants of STFTSynth were developed by selectively removing these components to assess their impact on model performance. The evaluation specifically focused on wire breakage STFT spectrograms.
\begin{table}[H]
    \centering
    \caption{Performance metrics of different STFTSynth variants on wire breakage STFT spectrograms.}
    \label{tab:ablation}
    
    \resizebox{0.6\textwidth}{!}{
    \scriptsize
    \renewcommand{\arraystretch}{1.5} 
    \setlength{\tabcolsep}{5pt}   
    \begin{tabularx}{0.53\linewidth}{@{}lllll@{}}
    \toprule
    \textbf{STFTSynth Variants} & \textbf{SSIM} & \textbf{PSNR} & \textbf{FID} & \textbf{TTE [s] \footnotemark} \\ 
    \toprule
    \textbf{Full Model} & 0.488 & 13.7310 & 0.210 & 15.50 \\  
        \textbf{No DRB} & 0.431 & 13.1517 & 0.260 & 15.00 \\  
        \textbf{No BiGRU} & 0.346 & 12.8312 & 0.330 & 2.50 \\  
        \textbf{No DRB \& BiGRU} & 0.367 & 13.4090 & 0.219 & 1.10 \\  
    \bottomrule
    \end{tabularx}
    }
    \caption*{\footnotesize \scriptsize{TTE: Training Time per Epoch (in seconds)}}
\end{table}
The results presented in Table \ref{tab:ablation} clearly illustrate the essential roles of DRB and BiGRU layers in enhancing STFTSynth’s performance for generating high-quality STFT spectrograms. The full STFTSynth model, incorporating both DRB and BiGRU, achieves the highest performance across all metrics, with the best SSIM (0.488), highest PSNR (13.731), and lowest FID (0.210). These values confirm its strong capability to generate realistic and structurally detailed spectrograms.

Removing the DRB layers from STFTSynth leads to a performance decline across all metrics, especially reflected by an increased (worsened) FID score, rising from 0.210 to 0.260. This demonstrates the importance of the DRB layers in capturing detailed spectral structures. However, removing the DRB has a minimal effect on computational cost, as indicated by the Training Time per Epoch (TTE), which remains nearly unchanged. The removal of BiGRU layers has a greater degradation in performance, with SSIM dropping to 0.346, PSNR reducing to 12.831, and the FID score increasing notably to 0.330. This substantial drop highlights BiGRU's essential role in modeling temporal dependencies, critical for accurately capturing realistic STFT spectrogram patterns. However, eliminating the BiGRU layers significantly improves computational efficiency, reducing the TTE to just 2.50 seconds per epoch. Thus, the absence of BiGRU introduces a clear trade-off between computational efficiency and the quality of generated spectrograms. The effectiveness of the BiGRU layers can be attributed to the temporal characteristics of the STFT spectrogram, which encodes signal evolution across both time and frequency. Although convolutional layers are effective at capturing local spatial features, they are intrinsically limited in modeling long-range temporal dependencies. In contrast, BiGRUs process sequences in both forward and backward directions, enabling the model to learn temporal dependencies from both past and future contexts. This bidirectional processing is particularly beneficial for vibration signals, where structural anomalies such as wire breakage may exhibit subtle patterns that evolve over time and are not strictly one-directional. This way, BiGRUs can enhance the generator’s ability to produce more realistic and temporally coherent spectrograms. 

Removing both DRB and BiGRU layers returns STFTSynth to its original WGAN-GP baseline. This variant performs slightly better than scenarios where only one of these components is removed, achieving an FID score of 0.219, lower than both the "No DRB" and "No BiGRU" variants. However, the SSIM (0.367) and PSNR (13.409) values remain substantially lower compared to the full STFTSynth, clearly emphasizing that the integration of both DRB and BiGRU significantly enhances the model's ability to produce realistic STFT spectrograms. This baseline variant, however, offers the fastest computational efficiency, achieving the lowest TTE of only 1.10 seconds per epoch.

Overall, the ablation study conclusively demonstrates that integrating both DRB and BiGRU layers into the STFTSynth architecture provides substantial improvements in the quality of generated STFT spectrograms. Despite increased computational costs, the integration of both layers is justified by the enhanced performance. 

\section{Limitations and Future Work}
\label{sec6}
Even though the proposed framework considers all generated samples uniformly during training, their practical value in SHM application is not uniform. For instance, rare, high-energy transient events such as wire breakage can benefit more from synthetic augmentation than operational or environmental noise signals as they are more difficult to reproduce experimentally. From a data-centric perspective, synthetic data is particularly useful in scenarios where datasets are highly imbalanced or when specific events are extremely rare, which is commonly the case for damage events such as prestressing wire breakage in SHM. In such situations, collecting additional real-world data cannot be possible or it might be costly. It is also important to note that simply increasing the amount of data does not always lead to better model performance. When synthetic samples do not introduce additional variability beyond what is already present in the real data, the benefits of augmentation may be saturated. Understanding how different proportions of synthetic data influence model performance is an important topic for future research, particularly in downstream SHM tasks. Besides, some learning models can perform reasonably well under rare and imbalanced data conditions, their results may still lack reliability and generalizability due to the limited availability of representative samples. These considerations highlight the need for data-centric strategies that carefully balance synthetic augmentation and real data acquisition, which we identify as a promising direction for future work.

Although the proposed GAN-based framework shows promising results, it remains constrained by the diversity and quality of the available real-world dataset. Synthetic generation can only reflect the variability present in the original samples. Expanding data collection across different structures and operational conditions is therefore essential, and future work should also investigate semi-supervised and self-supervised approaches to improve performance under limited labeling. In addition, the present study focuses on generating synthetic spectrograms without integrating them into a downstream classification task. Due to the rarity of real wire breakage events, building a statistically meaningful and balanced classification benchmark was not feasible in the current setting. Future studies should therefore evaluate the effect of GAN-augmented datasets on SHM classifiers in order to assess their practical utility more directly.

STFT was chosen for its fixed-size outputs and compatibility with GAN training, however, alternative representations such as wavelet scalograms may also be beneficial to capture highly non-stationary or overlapping events. Diffusion-based generative models also present a promising alternative, offering improved stability, sample diversity, and the potential to reconstruct time-domain signals for more direct physical validation. Finally, real-world monitoring often involves overlapping acoustic events, which were absent in this controlled dataset. Addressing this challenge will require integrating source separation techniques to separate concurrent signals, improving both data generation and downstream event detection.

\section{Conclusion}
\label{sec7}
This study proposed a GAN-based framework for representation-level synthesis of STFT spectrograms associated with rare acoustic events in SHM. A novel customized model, STFTSynth, was introduced to generate realistic STFT spectrograms of various structural events. By integrating Dilated Residual Blocks and Bidirectional GRU layers, the model enhances both spectral and temporal coherence of the generated data. STFTSynth was evaluated against established baselines—DCGAN, WGAN-GP, and LSGAN—using quantitative metrics (SSIM, PSNR, and FID) and qualitative visual assessments.

The results demonstrate that STFTSynth consistently achieves the highest SSIM and PSNR values, indicating its superior ability to generate high-quality spectrograms with minimal distortion. FID scores further confirm that STFTSynth produces the most realistic samples in the majority of cases, although LSGAN shows competitive performance for specific classes, such as the Trimmer event. In contrast, WGAN-GP consistently exhibited the weakest performance, with lower SSIM and PSNR values, highlighting its limitations in preserving fine-grained structural and visual details. Qualitative assessments align closely with the quantitative results, with STFTSynth-generated spectrograms visually resembling real-world patterns more accurately than those produced by other models. The findings validate STFTSynth as an effective representation-level generator of single-channel STFT spectrograms for rare acoustic events in SHM. The results demonstrate improved spectrogram fidelity and distributional similarity compared with the selected GAN baselines. The effect of the generated samples on downstream event classification, damage detection, or anomaly detection remains a direction for future work.

\section*{Acknowledgment}
The authors would like to express their sincere gratitude to Strada dei Parchi SpA for providing the opportunity and support necessary to conduct the experimental tests.

\section*{Competing Interests}
The authors declare no competing interests.

\section*{Data Availability}
The data used in this study are not publicly available due to confidentiality and infrastructure security restrictions associated with the monitored site. 

A limited subset of the data, or derived features, may be made available from the corresponding author upon reasonable request, subject to approval by the data owner.

\section*{Author Contributions}
S.F.: Conceptualization, Data Curation, Methodology, Software, Validation, Formal analysis, Writing---original draft \& editing.
Co-author M.I.: Methodology, Software, Formal analysis, Writing---original draft.
Co-author M. C.: Supervision, Investigation, Data curation, Writing---review \& editing.
Co-author E. C.: Supervision, Writing---review \& editing.
Co-author G. V.: Supervision, Writing---review \& editing.
All authors approved the final manuscript.

\section*{AI Statement}
Generative AI tools were used to improve the clarity and readability of the manuscript (e.g., grammar and style suggestions). All technical content, experimental design, results, and interpretations were produced and verified by the authors, who take full responsibility for the manuscript.

\bibliographystyle{elsarticle-num} 
\bibliography{references}

\end{document}